\documentclass[pdflatex,sn-mathphys-num]{sn-jnl}% Basic Springer Nature Reference Style/Chemistry Reference Style
\usepackage{graphicx}%
\usepackage{multirow}%
\usepackage{amsmath,amssymb,amsfonts}%
\usepackage{amsthm}%
\usepackage{mathrsfs}%
\usepackage[title]{appendix}%
\usepackage[dvipsnames]{xcolor}%
\usepackage{textcomp}%
\usepackage{manyfoot}%
\usepackage{booktabs}%
\usepackage{algorithm}%
\usepackage{algorithmicx}%
\usepackage{algpseudocode}%
\usepackage{listings}%

\usepackage{hyperref}
\usepackage{xspace}
\usepackage[inline]{enumitem}
\usepackage{booktabs}
\usepackage{makecell}
\usepackage{tcolorbox}
\usepackage{balance}
\usepackage{amsthm}
\usepackage{tikz}
\usepackage{pifont}% http://ctan.org/pkg/pifont
\usepackage{subcaption}
\usepackage{utfsym}
\usepackage[title]{appendix}
\hypersetup{
    colorlinks=true,
    linkcolor=Fuchsia,     % internal links (sections, refs)
    citecolor=Plum,     % citations
    urlcolor=Fuchsia       % external URLs
}

\newcommand{\newcheckmark}{\usym{1F5F8}}
\newcommand{\newcrossmark}{\scalebox{0.75}{\usym{2613}}}

\newcommand{\etal}{\textit{et al.}\xspace}

\newcommand{\modified}[1]{\textcolor{black}{#1}}

\newcommand{\manifestdef}{dependency specification\xspace}
\newcommand{\manifest}{dependency specification file\xspace}
\newcommand{\manifests}{dependency specification files\xspace}

\newcommand{\nppm}{seven\xspace}
\newcommand{\npinvitations}{59\xspace}

\newcommand{\npinterviews}{15\xspace}
\newcommand{\npdomains}{12\xspace}
\newcommand{\npavgexperience}{eight\xspace}
\newcommand{\npthemes}{seven\xspace}
\newcommand{\npsubthemes}{20\xspace}

\newcommand{\npinterviewreplicratio}{6/9\xspace}

\newcommand{\nbrecommendationsus}{two\xspace}
\newcommand{\nbrecommendationsdev}{three\xspace}
\newcommand{\nbrecommendations}{five\xspace}

\newcommand{\nptotalrepos}{4,859\xspace}
\newcommand{\nptotallock}{3,956\xspace}
\newcommand{\nptotallockbefore}{3,040\xspace}
\newcommand{\nptotallockratio}{81.4}

\newcommand{\nptotallockratiobefore}{62.6}
\newcommand{\nptotallocklater}{916}
\newcommand{\nptotallockratiolater}{18.8}

\newcommand{\npcargo}{1,089\xspace}
\newcommand{\npcargolock}{772\xspace}
\newcommand{\npcargolockratio}{70.9}
\newcommand{\npcargolockbefore}{596\xspace}%after 176
\newcommand{\npcargolockratiobefore}{54.7}

\newcommand{\nppoetry}{314\xspace}
\newcommand{\nppoetrylock}{263\xspace}
\newcommand{\nppoetrylockratio}{83.8}
\newcommand{\nppoetrylockbefore}{155\xspace}

\newcommand{\nppoetrylockratiobefore}{49.3}

\newcommand{\nppipenv}{29\xspace}
\newcommand{\nppipenvlock}{25\xspace}
\newcommand{\nppipenvlockratio}{86.2}
\newcommand{\nppipenvlockbefore}{16\xspace}
\newcommand{\nppipenvlockratiobefore}{55.2}

\newcommand{\npjsts}{1,916\xspace}

\newcommand{\npnpmlock}{1,021\xspace}
\newcommand{\npnpmlockratio}{53}
\newcommand{\npnpmlockbefore}{832\xspace} %after 189
\newcommand{\npnpmlockratiobefore}{43.4} 
\newcommand{\nppnpmlock}{688\xspace}
\newcommand{\nppnpmlockratio}{36}
\newcommand{\nppnpmlockbefore}{345\xspace}%after 343
\newcommand{\nppnpmlockratiobefore}{18}

\newcommand{\npgradle}{323\xspace}
\newcommand{\npgradlelock}{3\xspace}
\newcommand{\npgradlelockratio}{0.9}
\newcommand{\npgradlelockbefore}{1\xspace}
\newcommand{\npgradlelockratiobefore}{0.3}
\newcommand{\npgo}{1,188\xspace}
\newcommand{\npgolock}{1,184\xspace}

\newcommand{\npgolockratio}{99.7}
\newcommand{\npgolockbefore}{1,095\xspace} %89
\newcommand{\npgolockratiobefore}{92}

\newcommand{\sap}{P1\xspace}
\newcommand{\libre}{P2\xspace}
\newcommand{\uap}{P3\xspace}
\newcommand{\grammy}{P4\xspace}
\newcommand{\neocities}{P5\xspace}
\newcommand{\redlib}{P6\xspace}
\newcommand{\sniffnet}{P7\xspace}
\newcommand{\bevy}{P8\xspace}
\newcommand{\gramrs}{P9\xspace}
\newcommand{\wyze}{P10\xspace}
\newcommand{\patito}{P11\xspace}
\newcommand{\scrapli}{P12\xspace}
\newcommand{\gym}{P13\xspace}
\newcommand{\chatgpt}{P14\xspace}
\newcommand{\gomail}{P15\xspace}

\theoremstyle{definition}
\newtheorem{definition}{\textit{Definition}}

\definecolor{codegreen}{rgb}{0,0.6,0}
\definecolor{codegray}{rgb}{0.5,0.5,0.5}
\definecolor{codelightgray}{rgb}{0.85,0.85,0.85}
\definecolor{codepurple}{rgb}{0.58,0,0.82}
\definecolor{backcolour}{rgb}{245,245,245}

\lstdefinelanguage{json}{
    basicstyle=\ttfamily\scriptsize,
    numbers=left,
    numberstyle=\tiny\color{codegray},
    stepnumber=1,
    numbersep=5pt,
    showstringspaces=false,
    breaklines=true,
    frame=single,
    backgroundcolor=\color{backcolour},
    literate=
     *{:}{{{\color{red}{:}}}}{1}
      {,}{{{\color{red}{,}}}}{1}
      {\{}{{{\color{red}{\{}}}}{1}
      {\}}{{{\color{red}{\}}}}}{1}
      {[}{{{\color{red}{[}}}}{1}
      {]}{{{\color{red}{]}}}}{1},
}

\tcbset{
    mylistingstyle/.style={
        colback=backcolour,
        colframe=black!75!black, % Frame color
        boxrule=0.1pt,           % Border thickness
        arc=0pt,                 % No rounded corners
        boxsep=0pt,              % Internal padding
        top=0pt,                 % Remove top padding
        bottom=0pt               % Remove bottom padding
    }
}

\lstdefinestyle{mystyle}{ 
    backgroundcolor=\color{backcolour}, 
    commentstyle=\color{codegreen},
    keywordstyle=\color{magenta},
    numberstyle=\tiny\color{codegray},
    stringstyle=\color{codepurple},
    basicstyle=\ttfamily\scriptsize,
    breakatwhitespace=false,         
    breaklines=true,                 
    captionpos=b,                    
    keepspaces=true,                 
    numbers=left,                    
    numbersep=5pt,                  
    showspaces=false,                
    showstringspaces=false,
    showtabs=false,                  
    tabsize=4
}

\begin{document}

\title{The Design Space of Lockfiles Across Package Managers}

\author[1]{\fnm{Yogya} \sur{Gamage}}\email{yogya.gamage@umontreal.ca}

\author[2]{\fnm{Deepika} \sur{Tiwari}}\email{deepikat@kth.se}

\author[2]{\fnm{Martin} \sur{Monperrus}}\email{monperrus@kth.se}

\author[1]{\fnm{Benoit} \sur{Baudry}}\email{benoit.baudry@umontreal.ca}

\affil[1]{\orgname{Université de Montréal}, \orgaddress{\city{Montreal}, \country{Canada}}}

\affil[2]{ \orgname{KTH Royal Institute of Technology}, \orgaddress{\city{Stockholm}, \country{Sweden}}}


\abstract{Software developers reuse third-party packages that are hosted in package registries.
At build time, a package manager resolves and fetches the direct and indirect dependencies of a project.
Most package managers also generate a lockfile, which  records the exact set of resolved dependency versions. 
Lockfiles are used to reduce build times; to verify the integrity of resolved packages; and to support build reproducibility across environments and time.
Despite these beneficial features, developers often struggle with their maintenance, usage, and interpretation.
In this study, we unveil the major challenges related to lockfiles, such that future researchers and engineers can address them.

We perform the first comprehensive study of lockfiles across \nppm popular package managers, npm, pnpm, Cargo, Poetry, Pipenv, Gradle, and Go.
First, we highlight the wide variety of design decisions that package managers make, regarding the generation process as well as the content of lockfiles. 
Next, we conduct a qualitative analysis based on semi-structured interviews with \npinterviews developers.
We capture first-hand insights about the benefits that developers perceive in lockfiles, as well as the challenges they face to manage these files.
Following these observations, we make \nbrecommendations recommendations to further improve lockfiles, for a better developer experience.}

% \keywords{keyword1, Keyword2, Keyword3, Keyword4}



\maketitle

\section{Introduction}

% Background: software supply chain, dependency management, reproducibility, 
% dependency management problems
Software reuse is an established best practice in software engineering \cite{Krueger1992}.
Software developers reuse third-party libraries for commodity features such as logging, cryptography, arithmetic, or I/O.
Reuse helps developers save time, pool quality assurance efforts, and focus on the core, application specific features of their project \cite{mohagheghi2007quality}.
Yet, software reuse at scale introduces major engineering challenges \cite{Cox19}.
The use of third-party libraries, including the indirect ones, often results in large dependency trees.
Managing these increasingly complex dependency trees is hard because developers have to resolve compatible versions across the whole tree \cite{rausch2017empirical}, stay on top of updates to patch bugs and vulnerabilities \cite{ladisa2023sok}, and ensure that new changes do not break the build \cite{frank2024bg}.

% Package managers, spec file, dependency resolution, leading to lockfiles
Package managers are tools that assist developers in the management of reused libraries \cite{spinellis2012package}.
They automate and simplify the configuration, resolution, installation, and update of dependencies.
When using a package manager, developers specify the dependencies of a project in a \manifest.
\modified{The package manager then resolves compatible versions based on a combination of developer-specified constraints. Dependency resolution typically defaults to the latest available version if no version is specified.
Then, the package manager fetches resolved dependencies from a package registry, and installs them to further build the project.}
During this process, some package managers save the exact list of resolved dependencies and their versions into a special file: the \emph{lockfile} \cite{yu2024sbomgen}. 

% Motivation
Lockfiles are widely used throughout the software development lifecycle, supporting reproducible builds \cite{Lamb2022},  defending the software supply chain \cite{liu2022demystifynpm, Mohayeji2025dependabot}, and contributing to improved build performance \cite{aïdasso2025buildoptimizationsystematicliterature}. 
However, lockfiles also introduce additional complexity, and their maintenance requirements are not fully understood, a subject hardly touched by software engineering research.
Lockfiles have been mentioned as security best practices \cite{kabir2022securitybest}, important for software supply chain security \cite{williams25directions}, related to Software Bill of Materials (SBOM) generation  \cite{Bi2024, yu2024sbomgen}, and notable for new dependency management tools \cite{cleare2018gem}.
However, none of these works provide an in-depth analysis of lockfiles, why they exist and what they should contain.
To our knowledge, there is no systematic analysis and comparison of the different strategies in the field to generate and evolve lockfiles. 
There is no grounded evidence on how developers perceive the benefits and challenges of maintaining lockfiles.

% Methodology: comprehensive comparison of algorithms and features, then interviews
\modified{In this study, we address this critical gap in the software engineering literature by  conducting a comprehensive analysis of lockfiles across \nppm package managers.
These package managers serve the top programming languages of 2024 \cite{cass2024spectrum}: Javascript, Python, Java and Golang.
The considered package managers are:
npm and pnpm for Javascript, Cargo for Rust, Poetry and Pipenv for Python, Gradle for Java, and Go for Golang. 
% RQ1
We deeply analyze the implementation of these package managers to uncover how they generate, update, and structure lockfiles.
% RQ2
We then quantitatively evaluate the extent to which developers include lockfiles in version control by collecting GitHub projects and analyzing their lockfile commit rates.
% RQ3 and RQ4
To complement this analysis, we conduct interviews with \npinterviews developers from diverse backgrounds and experiences, who actively  contribute to open-source. 
Through these interviews, we gather valuable and grounded first-hand insights into the benefits developers perceive in using lockfiles, as well as the challenges they encounter when working with them.} 

% Key findings
Our paper is the first study that highlights how the functionalities and usage of lockfiles differ across package managers and ecosystems, resulting in varying levels of developer satisfaction. 
\modified{While some approaches, such as Go's strategy to always generate a lockfile by default, facilitates their adoption by developers, others, such as Gradle that requires specializing the build in order to have a lockfile, fails to make the lockfile usable.}
Our findings clearly show that a simple, principled approach to the lockfile lifecycle leads to a better developer experience. 
We also gather evidence of the challenges that developers face  with lockfiles, such as limited human readability, difficulty locking indirect dependencies, and delays in receiving dependency updates. 
Overall, developers generally trust lockfiles and view them as a valuable tool for improving the efficiency and security of dependency management. 
We conclude our study with \nbrecommendations recommendations to encourage the adoption of lockfiles in all projects.
Our findings are directly actionable by package manager users  and by package manager maintainers. Our results assist package manager users in understanding lockfiles and choosing a package manager that better suits their project requirements. Our findings  help package manager maintainers improve their lockfile feature, which in turn can positively impact a large community of developers. 
% list of contributions
In summary, our key contributions are, 
\begin{itemize}
    \item the first comprehensive comparison of lockfiles across the  \nppm top package managers, with a deep analysis of the key dimensions of the lockfile problem space: the lockfile lifecycle, and content of the generated lockfile.
    \item a qualitative assessment of the benefits and challenges of lockfiles, through \npinterviews developer interviews, demonstrating that developers well acquainted with lockfiles see essential value for addressing important software engineering challenges.
    \item a regimen of \nbrecommendationsus recommendations for package manager users and \nbrecommendationsdev recommendations for package manager developers to further improve lockfile usage for dependency management, to inform future research in the field as well as engineering of the future generation of package managers.
\end{itemize}

\section{Definitions and Context}

% https://drive.google.com/file/d/1mld2JlARarUBPn9RUShZbvW171qtW-Ti/view?usp=sharing

\begin{figure}
\centering
\includegraphics[width=0.85\linewidth]{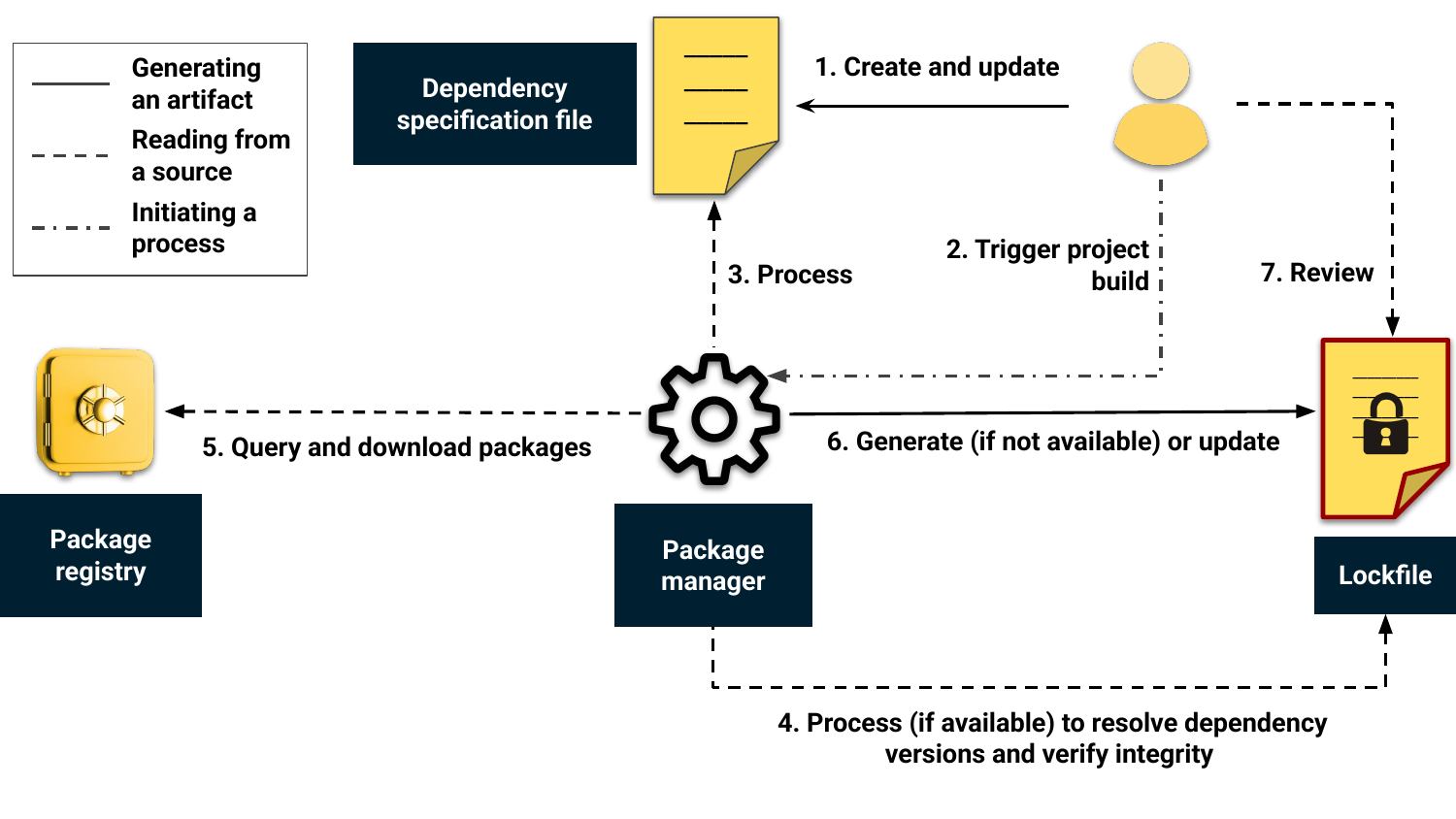}
\caption{A canonical workflow of interactions between a developer, the \manifest, the package manager, the lockfile, and the package registry. The order of events may slightly vary depending on the exact command executed by the developer in step 2.}
%The boxes next to the \manifest and lockfile icons contain excerpts from an npm \manifest and lockfile in the asyncapi/modelina project.
\label{fig:package-manager-diagram}
\end{figure}

Software developers who wish to reuse third-party packages, typically specify the configurations for these packages in a \emph{\manifest}.
Then, developers can rely on a \emph{package manager} to help them install and maintain reusable code packages efficiently. 
When developers invoke an action that involves resolving the dependencies, the package manager parses the \manifest, queries the default or specified alternative \emph{registries}, determines the required package versions, and downloads them. During this process, it may also record the resolved direct and indirect package versions in a \emph{lockfile}. Developers can review this file to inspect the exact versions selected by the package manager, and may commit it to version control for reproducible builds, improved security, and efficiency. In subsequent project builds the package manager can refer to the lockfile to quickly resolve required dependency versions, perform integrity checks against the downloaded packages and ensure deterministic project builds. This process is illustrated in \autoref{fig:package-manager-diagram}. In the remainder of this section, we define these different key concepts of the package manager, which will be used throughout the paper.

\subsection{Definitions}

\begin{definition}
\label{Package Manager}
\modified{A \textbf{package manager} is a tool that automates code reuse by ensuring that all third-party packages required by the project, directly or indirectly, are downloaded with the correct versions.} Most popular programming languages have one or more package managers. \modified{For example, Cargo is the package manager for Rust.} Package managers typically provide developers with a command line interface and sometimes a graphical user interface to interact with the package registry. 
\end{definition}

\begin{definition}
\label{Package Registry}
A \textbf{package registry} is as a central platform for sharing code across multiple projects. Depending on the registry, developers can upload packaged binaries or packaged source code, allowing others to download the code to be reused within their own projects. Popular programming languages have official package registries. \modified{Crates.io is the package registry for Rust and developers can distribute their packaged Rust source code through a central registry at Crates.io.} 
\end{definition}

\begin{definition}
\label{manifest}
A \textbf{\manifestdef} is a file that specifies the packages that a software project directly depends on, i.e., it specifies the \emph{direct} dependencies of the project.
\modified{For example, the \manifest for Cargo is a file called Cargo.toml.}
\modified{For each dependency, developers can either declare a specific version, a practice commonly known as pinning, specify a range of versions, which usually resolves to the latest version within that range, or specify no version, which means resolving to the available latest version.}
The package manager takes the \manifest as input to build the project.
The \manifest also includes additional information related to dependency management, such as metadata about the project or exclusion of certain dependencies.
The \manifest is primarily maintained by developers, and also by automated dependency bots such as Dependabot.
\end{definition}

\begin{definition}
\label{lockfile}
A \textbf{lockfile} is a file that is automatically generated by the package manager and that is not meant to be edited by developers. 
\modified{Each package manager uses its own file naming convention for the lockfile. For example, in Cargo, the file is named Cargo.lock.}
The lockfile records the resolved versions of all direct and indirect dependencies that need to be included in a project. Some package managers may include extra details, such as the hash of the \manifest content or the package manager version. 
% who uses, for what
Package managers use lockfiles when (re)building projects, such as for fetching the exact versions of dependencies required, and verifying their integrity. 
Lockfiles may also be read by developers during code reviews to audit resolved dependencies, and updated by automated dependency bots during dependency updates.
\end{definition}

\begin{table*}
\small
\caption{The \nppm package managers analyzed in this study, including their associated programming languages, registries, dependency specifications, and lockfiles.}
\label{tab:ecosystems}
\centering

%\rowcolors{2}{gray!10}{white}
\begin{tabular}{rrrrrr}
\toprule
Package manager & Language & \makecell[r]{Package registry} & \makecell[r]{Dependency \\ specification \\ file} & Lockfile \\
\hline

npm CLI & JavaScript & \href{https://www.npmjs.com/}{npm.com} & package.json & package-lock.json \\
pnpm & JavaScript & \href{https://www.npmjs.com/}{npm.com} & package.json & pnpm-lock.yaml \\
Cargo & Rust & \href{https://crates.io/}{crates.io} & Cargo.toml & Cargo.lock \\
Poetry & Python & \href{https://pypi.org/}{PyPi.org} & pyproject.toml & poetry.lock \\
Pipenv & Python & \href{https://pypi.org/}{PyPi.org} & Pipfile & Pipfile.lock \\
Gradle & Java & \href{https://central.sonatype.com/}{\makecell[r]{central.sonatype}} & build.gradle & gradle.lockfile \\
% \makecell[r]{\href{https://proxy.golang.org/}{Go Module Mirror}, \\ \href{https://index.golang.org/}{Index}, \\ \href{https://sum.golang.org/}{Checksum Database}}
Go & Go & \href{https:github.com/}{github.com} & go.mod & go.mod \& go.sum \\
\bottomrule
\end{tabular}

\end{table*}

\subsection{Considered Package Managers}

\modified{We select \nppm popular package managers for our study based on the popularity of programming languages according to IEEE Spectrum in 2024  \cite{cass2024spectrum}}. From the ten most popular languages, we identify those with widely used package managers that support lockfiles: JavaScript, TypeScript, Python, Java, and Go. We add Rust due to its significant upward trend. \modified{For each language, we select the most commonly used package manager with official support for lockfiles: npm for JavaScript/TypeScript, Poetry for Python, Cargo for Rust, Gradle for Java, and Go's built-in toolchain.} When a language has multiple package managers, we  add a second one: pnpm for JavaScript and Pipenv for Python. \modified{We use GitHub as the source for identifying the most widely used package managers.}\autoref{tab:ecosystems} outlines the \nppm package managers analyzed in this study, along with their associated languages, registries, \manifests, and lockfiles. 

% example NPM
For instance, npm CLI is the default package manager for JavaScript, and the npm registry is the default package registry. Library developers can upload their JavaScript code to the npmjs.com registry. 
The \manifest and lockfile for npm CLI are in JSON format and are named \texttt{package.json} and \texttt{package-lock.json}, respectively. Excerpts from a package.json and its corresponding package-lock.json are given in \autoref{lst:example-npm-manifest} and \autoref{lst:example-npm-lockfile}, respectively. \modified{These excerpts show the basic types of metadata present in the lockfiles, such as the project name, project version, and list of dependencies. Note that the lockfile includes the exact version resolved for each dependency under the field \textit{version}, along with a link to the source and its checksum under the fields \textit{resolved} and \textit{integrity}.}

In the case of Go, there is no centralized package registry. Instead, Go modules are downloaded from version control registries such as GitHub. Module versions are tracked using an index that feeds available new versions. The integrity verification of Go packages is done based on a checksum database.
Moreover, in Go, the \texttt{go.mod} file works both as the \manifest and the lockfile. An additional file, named \texttt{go.sum}, is generated by the package manager, to record the checksums of the downloaded modules. The \texttt{go.mod} file is not meant to be edited manually by developers but should be modified using specific commands. \autoref{lst:example-go-mod} and \autoref{lst:example-go-sum} show an example \textit{go.mod} file and a \textit{go.sum} file. \modified{In this example, the \textit{go.mod} file contains the module name, Go version, and the dependencies, whereas the \textit{go.sum} includes the checksums of the resolved dependencies.}

\modified{\autoref{tab:lockfileRepos} provides the list of selected package manager versions, including the number of lines of code in the analyzed version, the number of available releases up to that version, and the date of the first release, which together indicate the complexity and maturity of these package managers.} The release counts and first release dates are obtained from the official releases page of each package manager, except for pnpm, where they are based on information from the npm registry. 
% These values may vary across sources.
% The subsequent analyses in this paper are based on the package manager versions listed in column 2 of \autoref{tab:lockfileRepos}. 

\begin{table*}
\small
\caption{Details about the exact versions of the \nppm package managers under study.}
\label{tab:lockfileRepos}
\centering

%\rowcolors{2}{gray!10}{white}
\begin{tabular}{rrrrr}
\toprule
\makecell[l]{Package \\ manager} & Analyzed Version & Lines of code & \makecell[l]{Versions} & \makecell[l]{First \\ release}\\
\hline
npm CLI & \href{https://github.com/npm/cli/tree/v11.1.0}{npm/cli} (11.1.0) & 459k (JavaScript) & 557 & 12/2013\\

pnpm & \href{https://github.com/pnpm/pnpm/tree/v10.1.0}{pnpm/pnpm} (10.1.0) & 112k (TypeScript) & 1,103 & 06/2017\\

Cargo & \href{https://github.com/rust-lang/cargo/tree/0.85.0}{rust-lang/cargo} (0.85.0 ) & 210k (Rust) & 119 & 10/2018\\

Poetry & \href{https://github.com/python-poetry/poetry/tree/2.0.1}{python-poetry/poetry} (2.0.1) & 46k (Python) & 88 & 03/2018\\

Pipenv & \href{https://github.com/pypa/pipenv/tree/v2024.4.1}{pypa/pipenv} (2024.4.1) & 140k (Python) & 396 & 06/2018\\

Gradle & \href{https://github.com/gradle/gradle/tree/v8.12.1}{gradle/gradle} (8.12.1) & \makecell[r]{784k (Groovy) \\ 531k (Java) \\ 367k (Kotlin)} & 148 & 07/2009\\

Go & \href{https://github.com/golang/go/tree/go1.24.0}{golang/go} (1.24.0) & \makecell[r]{2,052k (Go)} & 243 & 03/2012\\

\bottomrule
\end{tabular}
    
\end{table*}

\section{Comprehensive Comparison of Lockfile Design Decisions}

Across ecosystems, lockfiles serve different purposes such as reproducible builds, package integrity verification, or performance improvement for project builds.
Accordingly, different package managers make distinct design decisions when implementing the lockfile feature, depending on their priorities. 
It is our intuition that the different approaches to implementing lockfiles influence how developers use lockfiles. In this section, we systematically map the differences between the main lockfile designs, particularly the content of lockfiles and their lifecycles across various package managers. We then conduct a quantitative analysis of the number of projects that include lockfiles in version control within each ecosystem.

\subsection{Research Questions}
\newcommand\rqLifecycle{How do package managers generate, update, and structure lockfiles?\xspace}

\newcommand\rqPatterns{To what extent do projects include lockfiles in version control?\xspace}

%%%%%%%%%%%%%%%%%%%%%%%%%%%%%%%%%%%%%%%%%%%%%%%%%%%

\begin{enumerate}[label=RQ\arabic*:, ref=RQ\arabic*]

    \item \label{rq:lifecycle}\textbf{\rqLifecycle}
    
    Lockfiles have been independently designed and implemented across ecosystems, with some inspiration flowing from one to another.
    Lockfiles must include relevant, useful information without excessive verbosity, in order to keep both efficient project builds and high developer usability. This RQ highlights the diversity among package managers with respect to the generation, enforcement, and structure of lockfiles. 
    Our study helps developers understand and adopt best practices for lockfile generation and management.
    %It may also be useful for developers who are getting started building projects within an ecosystem, or migrating from one package manager to another.

    \item \label{rq:usage}\textbf{\rqPatterns}

    In this RQ, we quantitatively evaluate how developers of different ecosystems include lockfiles in version control. Including lockfiles in version control systems is important to achieve all benefits lockfiles have to offer, including deterministic builds \cite{Goswami2020npmreproducible}. 
    We base our evaluation on lockfile commit data collected from public GitHub projects created within the past 5 years (from 2019 to 2024). 
\end{enumerate}

\begin{figure}[ht]
\centering
\begin{minipage}[t]{0.45\textwidth}
    \begin{lstlisting}[language={json}, caption={An excerpt from the specification file (package.json) of the project asyncapi/modelina}, label=lst:example-npm-manifest]
{
  "name": "@asyncapi/modelina",
  "version": "3.8.0", 
  "description": "Library for generating data models based on inputs such as AsyncAPI, OpenAPI, or JSON Schema documents",
...
"dependencies": {
    "@apidevtools/json-schema-ref-parser": "^11.1.0",
    "@apidevtools/swagger-parser": "^10.1.0",
    "@asyncapi/multi-parser": "^2.1.1",
    "@asyncapi/parser": "^3.1.0",
...
\end{lstlisting}
\end{minipage}%
\hfill
\begin{minipage}[t]{0.45\textwidth}
    \begin{lstlisting}[language={json}, caption={An excerpt from the lockfile (package-lock.json) of the project asyncapi/modelina}, label=lst:example-npm-lockfile]
{
  "name": "@asyncapi/modelina",
  "version": "3.8.0",
  "lockfileVersion": 3,
...
"node_modules/@apidevtools/json-schema-ref-parser": {
      "version": "11.1.1",
      "resolved": "https://registry.npmjs.org/@apidevtools/...",
      "integrity": "sha512-WZyqzPLrWctQB8qB8uGivvzWg...",
      "dependencies": {
        "@jsdevtools/ono": "^7.1.3",
...
\end{lstlisting}
\end{minipage}
\end{figure}

\begin{figure}[ht]
\centering
\begin{minipage}[t]{0.45\textwidth}
    \begin{lstlisting}[language={}, caption={An excerpt from the \manifest (go.mod) of the project piplabs/story}, label=lst:example-go-mod, frame=single]
module github.com/piplabs/story

go 1.22.11

require (
	cosmossdk.io/api v0.7.5
	// NOTE(https://github.com/cosmos/rosetta/issues/76): Rosetta requires cosmossdk.io/core v0.12.0 erroneously ...
	cosmossdk.io/core v0.12.0
	cosmossdk.io/depinject v1.0.0
	cosmossdk.io/errors v1.0.1 // indirect
...
\end{lstlisting}
\end{minipage}%
\hfill
\begin{minipage}[t]{0.45\textwidth}
    \begin{lstlisting}[language={}, caption={An excerpt from the lockfile  (go.sum) of the project piplabs/story}, label=lst:example-go-sum, frame=single]
buf.build/gen/go/bufbuild/bufplugin/protocolbuffers/go v1.36.3-... h1:dS5ier+mttGuW+lRLP/eC1CKJm2Rg3rKWy9Iy0hroIU=
buf.build/gen/go/bufbuild/bufplugin/protocolbuffers/go v1.36.3-.../go.mod h1:MYDFm9IHRP085R5Bis68mLc0mI...
buf.build/gen/go/bufbuild/protovalidate/protocolbuffers/go v1.36.3-...h1:cQZXKoQ+eB0kykzfJe80RP3nc+3PWbbBrUBm8XNYAQY=

...
\end{lstlisting}
\end{minipage}
\end{figure}

\subsection{Methodology for RQ1}

To answer the RQ1, we analyze the implementations of the lockfile feature in the source code of package managers, and we review their documentation.

% source code analysis
\textbf{Source code analysis}: To understand the workflows of lockfile-related tasks for each package manager, we analyze their implementation within their source code repository. 
We begin by locating the main function that is invoked after executing an install or build command. We then analyze the functions called by this main function, manually tracing their execution flow. We further search the project using regular expressions for the terms:  \textit{lockfile}, \textit{locked}, \textit{frozen}, and \textit{dependency resolution}. \modified{We use the term \textit{dependency resolution} as an umbrella term because lockfile related functions are generally tied to the dependency resolution processes. We also look for the terms \textit{locked} and \textit{frozen} as they are the options that users can specify to enable lockfile based functions in many package managers.}
The sections of source code, analyzed for each package manager are summarized in \autoref{tab:lockfileExperiments}.
For instance, in the case of npm, we study the repository {\textit{npm/cli}}, focusing on the source code for {\texttt{install}} and {\texttt{ci}} actions, dependency tree resolution, and lockfile validation.

% documentation
\textbf{Documentation}: In addition to analyzing the source code of the \nppm package managers, we review their official documentation.
We focus on sections related to dependency resolution actions such as install or build, and study the available configurations or flags that can be used with these actions.
We search for sections covering lockfiles, including lockfile-related recommendations and comparisons between lockfiles and \manifests.
%In Cargo documentation, we also explore the sections that distinguish the Cargo lockfile from the \manifest (cargo.toml).
For example, as given in \autoref{tab:lockfileExperiments}, for Poetry, we go through the section \textit{installing dependencies} that explain lockfile usage, including installing without poetry.lock, installing with poetry.lock, and committing poetry.lock to version control.
Similarly, for Pipenv, we analyze the \textit{commands} and \textit{Pipfile \& Pipfile.lock} sections that cover general lockfile recommendations, examples, and security features.
\modified{ 
Two authors of this paper conduct this analysis, and a third one contributes in case of a disagreement. Together, they decide on the features to analyze after an initial review of all repositories and documentation. In the second run, the two authors evaluate each package manager individually against the selected features. When the two authors do not agree, they refer to the opinion of the third author. In the end, they reach full agreement, with a documented pointer to the code  and documentation that describe each feature of interest.} 

%The list of features and behaviors we look for within the source code and documentations is decided upfront based on the \todo{ask about this}. The list of files and documentation sections selected for review is limited due to the careful selection of keywords.

\begin{table*}
\small
\caption{Lockfile features for the \nppm package managers under study. We study their source code and documentations.}
\label{tab:lockfileExperiments}
\centering

%\rowcolors{2}{gray!10}{white}
\begin{tabular}{lrr}
\toprule
\makecell[l]{Package \\ manager} & Sections of source code & \makecell[r]{Sections of \\ documentation}\\
\hline
npm CLI & 
\makecell[r]{\href{https://github.com/npm/cli/blob/latest/lib/commands/install.js}{install} and \href{https://github.com/npm/cli/blob/latest/lib/commands/ci.js}{ci} actions, \\ \href{https://github.com/npm/cli/tree/latest/workspaces/arborist}{dependency tree resolution} \\ (within \texttt{arborist} module), \\ \href{https://github.com/npm/cli/blob/latest/lib/utils/validate-lockfile.js}{lockfile validation}} & \makecell[r]{\href{https://docs.npmjs.com/cli/v11/commands}{CLI commands}} \\
\hline
pnpm & \makecell[r]{\href{https://github.com/pnpm/pnpm/tree/main/lockfile}{lockfile module}, \\ \href{https://github.com/pnpm/pnpm/tree/main/pkg-manager/core/src/install}{install action}} & \makecell[l]{\href{https://pnpm.io/cli/install}{\makecell[r]{CLI commands - \\ Manage dependencies}}} \\
\hline
Cargo & \makecell[r]{\href{https://github.com/rust-lang/cargo/blob/master/src/bin/cargo/commands/install.rs}{install action}, \\ \href{https://github.com/rust-lang/cargo/blob/master/src/cargo/ops/lockfile.rs}{lockfile functions}} & \makecell[r]{\href{https://doc.rust-lang.org/cargo/commands/build-commands.html}{Build commands}, \\ \href{https://doc.rust-lang.org/cargo/commands/manifest-commands.html}{manifest commands}, \\ \href{https://doc.rust-lang.org/cargo/guide/cargo-toml-vs-cargo-lock.html}{Cargo.toml vs Cargo.lock}} \\
\hline
Poetry & \makecell[r]{\href{https://github.com/python-poetry/poetry/blob/main/src/poetry/installation/installer.py}{installer class}, \\ \href{https://github.com/python-poetry/poetry/blob/main/src/poetry/packages/locker.py}{locker class}} & \href{https://python-poetry.org/docs/basic-usage/#installing-dependencies}{Installing dependencies} \\
\hline
Pipenv & \makecell[r]{\href{https://github.com/pypa/pipenv/blob/main/pipenv/cli/command.py}{command.py},  \href{https://github.com/pypa/pipenv/blob/main/pipenv/routines/install.py}{install.py}, \\ \href{https://github.com/pypa/pipenv/blob/main/pipenv/utils/locking.py}{locking.py},  \href{https://github.com/pypa/pipenv/blob/main/pipenv/routines/lock.py}{lock.py}} 
& \makecell[r]{\href{https://pipenv.pypa.io/en/latest/commands.html}{Pipenv commands,} \\ 
\href{https://pipenv.pypa.io/en/latest/pipfile.html}{Pipfile \& Pipfile.lock}} \\
\hline
Gradle & \makecell[r]
{\href{https://github.com/gradle/gradle/tree/master/platforms/software/dependency-management/src/main/java/org/gradle/internal/locking}{locking package}, \\
\href{https://github.com/gradle/gradle/tree/master/platforms/software/dependency-management/src/main/java/org/gradle/api/internal/artifacts/dsl/dependencies}{dependencies package}} & \makecell[r]{\href{https://docs.gradle.org/current/userguide/dependency_locking.html}{Locking versions}} \\
\hline
Go & \makecell[r]
{\href{https://github.com/golang/go/tree/go1.24.0/src/cmd/go/internal/modcmd}{modcmd}, \href{https://github.com/golang/go/tree/go1.24.0/src/cmd/go/internal/modload}{modload}, \\ 
\href{https://github.com/golang/go/tree/go1.24.0/src/cmd/go/internal/modfetch}{modfetch},
\href{https://github.com/golang/go/tree/go1.24.0/src/cmd/go/internal/modget}{modget},
\href{https://github.com/golang/go/tree/go1.24.0/src/cmd/go/internal/mvs}{mvs}} & \makecell[r]{\href{https://go.dev/doc/modules/managing-dependencies}{Managing dependencies} \\
\href{https://go.dev/ref/mod }{Go Modules Reference} \\
\href{https://go.dev/doc/modules/gomod-ref}{go.mod file reference}}
\\

\bottomrule
\end{tabular}
    
\end{table*}

\subsection{Results for RQ1: \textbf{\rqLifecycle}}

In this RQ, we first look at the fundamental differences in the lockfile content among the \nppm selected package managers introduced in \autoref{tab:ecosystems}. Next we discuss the key differences in lockfile lifecycles including lockfile generation, updates and enforcement.

\subsubsection{Lockfile Content}

Package managers may include various information in lockfiles, dependencies, checksums, and additional details such as licenses. 
\autoref{fig:example-lockfiles} illustrates excerpts from lockfiles across the selected \nppm package managers. Each of them contains an example representation of one dependency where the dependency versions, source links, checksums, and indirect dependencies are highlighted in yellow, blue, green, and pink, respectively.

% essential info
\autoref{tab:lockfileContent} presents the most common types of information recorded in lockfiles across different package managers. 

\begin{table*}
\small
\caption{The most common types of information included in lockfiles}
\label{tab:lockfileContent}
\centering

%\rowcolors{2}{gray!10}{white}
    \begin{tabular}{rrrrrr}
    \toprule
    \makecell[r]{Package \\ manager} & \makecell[r]{Resolved \\ package \\ versions} & \makecell[r]{Package \\ checksums} & \makecell[r]{Package \\ source} & \makecell[r]{Direct/\\ indirect \\ difference} & \makecell[r]{Additional \\ metadata}\\
    \toprule
    npm CLI & \newcheckmark & \newcheckmark & \newcheckmark & \newcheckmark & \makecell[r]{Indirect dependencies \\ (for each dependency), \\ Dependency types, \\ OS, engine, \\ CPU, requirements, \\ Licenses, \\ Funding information \\ }\\ \hline
    pnpm & \newcheckmark & \newcheckmark & \newcrossmark & \newcheckmark & \makecell[r]{Indirect dependencies \\ (under each dependency), \\ OS, engine, \\ CPU requirements} \\ \hline
    Cargo & \newcheckmark & \newcheckmark & \newcheckmark & \newcheckmark & \makecell[r]{Indirect dependencies \\ (for each dependency), \\ Language, OS, \\ engine requirements} \\ \hline
    Poetry & \newcheckmark & \newcheckmark & \newcrossmark & \newcheckmark & \makecell[r]{Indirect dependencies \\ (for each dependency), \\ Language requirements, \\ Descriptions} \\ \hline
    Pipenv & \newcheckmark & \newcheckmark & \newcrossmark & \newcrossmark & Language requirements \\ \hline
    Gradle & \newcheckmark & \newcrossmark & \newcrossmark & \newcrossmark & Dependency scopes \\ \hline
    Go & \newcheckmark & \makecell[r]{\newcheckmark \\ (in \texttt{go.sum})} & \newcheckmark & \newcheckmark & - \\
    \bottomrule
    \end{tabular}

\end{table*}

\textbf{Resolved package versions and checksums}
All of the studied package managers include resolved package versions in lockfiles.
Checksums are also included in all lockfiles except in the Gradle lockfile. \modified{Gradle only records the groupId (the namespace of the project), artifactId (the project name), and version (the project version) of the resolved dependencies, which is a significant limitation.}

\textbf{Link to source}
NPM and Cargo include a provenance link to the source of the downloaded package. \modified{Having a link to the source is useful for verifying that no tampering has occurred during the download from the package registry.}
In terms of naming, npm CLI uses the field \texttt{resolved} for source links, while Cargo uses \texttt{source}.
In Go, the source can be inferred from the module name declaration itself, as we see from the link to the GitHub repository for the module \texttt{text} in \autoref{lst:example-other-lockfile}.
Gradle, pnpm, Poetry, and Pipenv do not include a direct link to the source of the downloaded packages.

\textbf{Indirect dependencies}
% All lockfiles except Cargo lockfiles include indirect dependencies. - this is not the case (explained below under version ranges)
When recording indirect dependencies, lockfiles may keep the tree structure or flatten the list.
Pipenv, Gradle, and Go, do not present indirect dependencies in a nested way.
Go distinguishes indirect dependencies using the comment \texttt{//indirect}, whereas Pipenv and Gradle do not explicitly separate them from direct dependencies.
npm, pnpm, Cargo, and Poetry provide a tree structure of depth 1 (as highlighted in pink in \autoref{fig:example-lockfiles}).

\textbf{Version ranges}
Cargo omits the version ranges of indirect dependencies (when listed under each parent dependency) unless multiple versions of the same dependency are required. When a version is included, Cargo specifies the exact resolved version. In contrast, npm, pnpm, and Poetry include version ranges for indirect dependencies under a parent dependency, reflecting the same constraints set in the \manifests. In addition to version ranges, it is valuable for developers to have straightforward access to the exact resolved versions.

% Each indirect dependency is also included as a separate entry within the lockfile with its own metadata.

% \textbf{version range}   \todo{unclear: does it include both resolved and ranges or only range?} \todo{in table split column revolsed versions and checksums in two: direct and indirect. adapt the text accordingly. one point for direct/ one point for indirect.}

\textbf{Additional metadata}
Some package managers include additional details such as language requirements (e.g. Poetry, npm), operating systems (e.g. npm, Cargo) or dependency scope (e.g. npm, Gradle).
Moreover, npm CLI includes in the lockfile extra metadata such as license details and funding information, directly copied from package.json. \modified{One can note that not all of this metadata is included in the examples of \autoref{fig:example-lockfiles}. This is because they show excerpts from real-world project lockfiles, where this information is not available in the \manifest, hence absent in the lockfiles as well.}
Although recording as much information as possible for each dependency may seem a good choice, it also makes the lockfiles lengthier and harder to be reviewed by human developers. 

%\begin{tcolorbox}[boxrule=1pt,arc=.3em, left=4pt, right=4pt] 
% 2 sentences
\textbf{Summary of lockfile content}
All the package managers in this study record both direct and indirect dependencies in lockfiles. All except Gradle also include dependency checksums.
% This information is essential for verifying the integrity of a build or for reproducing a build.
However, among package managers, there are notable differences in the additional metadata included and the way this information is structured, which impact the compactness and readability of lockfiles.

\begin{figure}[H]
\tiny
\centering
% 25-02-24-10-51
% First row (2 lockfiles)
\begin{minipage}[t]{0.48\textwidth}
\begin{tcolorbox}[mylistingstyle]
\begin{lstlisting}[basicstyle=\footnotesize\ttfamily, escapechar=!, language={}, caption={\protect\small \texttt{package-lock.json} of the project asyncapi/modelina}, label=lst:example-npm-lockfile-2]
"node_modules/@eslint/eslintrc": {
!\colorbox{yellow}{"version": "2.1.4"}!,
!\colorbox{cyan}{"resolved":"https\\:registry.npmjs"}!,
!\colorbox{green}{"integrity":"sha512-269Zhnju8..."}!, 
"dev": true,
!\colorbox{pink}{
\parbox{0.87\textwidth}{
"dependencies": \{ \\
  "ajv": "\textasciicircum6.12.4", \\
\},
}}!
"engines": {
    "node": "^12.22.0"
 },
"funding": {
  "url": ""
 }
}
\end{lstlisting}
\end{tcolorbox}
\end{minipage}
\hfill
\begin{minipage}[t]{0.48\textwidth}
\begin{tcolorbox}[mylistingstyle]
\begin{lstlisting}[basicstyle=\footnotesize\ttfamily, escapechar=!, language={}, caption={\protect\small \texttt{pnpm-lock.yaml} of the project csfive/composing-programs-zh}, label=lst:example-pnpm-lock]
'@vitejs/plugin-vue@!\colorbox{yellow}{5.0.4}!':
!\colorbox{green}{resolution:{integrity:sha512-WS3R45...}}!
engines: {node: ^18.0.0 || >=20.0.0}
!\colorbox{pink}{\parbox{0.85\textwidth}{
peerDependencies: \\
  vite: \textasciicircum5.0.0 \\
}}!
\end{lstlisting}
\end{tcolorbox}
\end{minipage}

\vspace{0.1cm}

% Second row (3 lockfiles)
\begin{minipage}[t]{0.32\textwidth}
\begin{tcolorbox}[mylistingstyle]
\begin{lstlisting}[basicstyle=\footnotesize\ttfamily, escapechar=!, language={}, caption={\protect\small \texttt{Cargo.lock} of the project leptos-rs/cargo-leptos}, label=lst:example-cargo-lockfile]
[[package]]
name = "Inflector"
!\colorbox{yellow}{{version = "0.11.4"}}!
!\colorbox{cyan}{
\parbox{0.95\textwidth}{source = "registry+https://..."}}!
!\colorbox{green}{{checksum="fe438t67y6..."}}!
!\colorbox{pink}{
\parbox{0.95\textwidth}{
dependencies = [
 "lazy\_static"
]}
}!
\end{lstlisting}
\end{tcolorbox}
\end{minipage}
\hfill
\begin{minipage}[t]{0.32\textwidth}
\begin{tcolorbox}[mylistingstyle]
\begin{lstlisting}[basicstyle=\footnotesize\ttfamily, escapechar=!, language={}, caption={\protect\small \texttt{poetry.lock} of the project mingrammer/diagrams}, label=lst:example-poetry-lockfile]
[[package]]
name = "jinja2"
!\colorbox{yellow}{{version = "3.1.4"}}!
description = ""
optional = false
python-versions =">=3.7"
files =
[{file =".tar.gz", 
!\colorbox{green}{hash="sha:25646778..."}!}]
!\colorbox{pink}{
\parbox{1.03\textwidth}{
[package.dependencies] \\
MarkupSafe = ">=2.0"
}}!
\end{lstlisting}
\end{tcolorbox}
\end{minipage}
\hfill
\begin{minipage}[t]{0.32\textwidth}
\begin{tcolorbox}[mylistingstyle]
\begin{lstlisting}[basicstyle=\footnotesize\ttfamily, escapechar=!, language={}, caption={\protect\small \texttt{Pipfile.lock} of the project Drazzilb08/daps}, label=lst:example-pipfile-lock]
"plexapi": {
!\colorbox{green}{"hashes":["sha=256..."]}!,
"index": "pypi",
"markers": "python_version>='3.9'",
!\colorbox{yellow}{"version": "==4.16.1"}!
},
\end{lstlisting}
\end{tcolorbox}
\end{minipage}

\vspace{0.1cm}

% Third row (2 lockfiles)
\begin{minipage}[t]{0.48\textwidth}
\begin{tcolorbox}[mylistingstyle]
\begin{lstlisting}[basicstyle=\footnotesize\ttfamily, escapechar=!, language={}, caption={\protect\small \texttt{gradle.lockfile} of the project datastax/fallout}, label=lst:example-gradle-lockfile]
com.github.jknack:handlebars:!\colorbox{yellow}{4.3.1}!=testCompileClasspath,testRuntimeClasspath
\end{lstlisting}
\end{tcolorbox}
\end{minipage}
\hfill
\begin{minipage}[t]{0.48\textwidth}
\begin{tcolorbox}[mylistingstyle]
\begin{lstlisting}[basicstyle=\footnotesize\ttfamily, escapechar=!, language={}, caption={\protect\small \texttt{go.mod} and \texttt{go.sum} (below the dashed line) of the project piplabs/story}, label=lst:example-other-lockfile]
require (
  !\colorbox{cyan}{github.com/kr/text}! !\colorbox{yellow}{v0.2.0}! //indirect
)
-----------------------------
github.com/kr/text v0.2.0 h1:!\colorbox{green}{5Nx0Ya0ZqY2ygV366Qz...}!
github.com/kr/text v0.2.0/go.mod h1:!\colorbox{green}{eLer722TekiGuMkidMxCpM0...}!
\end{lstlisting}
\end{tcolorbox}
\end{minipage}

\caption{\small Example dependency representations extracted from lockfiles (or a mod file) across \nppm package managers. \modified{Each listing contains one extracted record corresponding to a direct dependency in a lockfile.} Dependency versions are highlighted in \colorbox{yellow}{yellow}, source links in \colorbox{cyan}{blue}, checksums in \colorbox{green}{green}, and indirect dependencies in \colorbox{pink}{pink}. Some details are omitted for brevity and clarity.}
\label{fig:example-lockfiles}
\end{figure}

%\end{tcolorbox}

\subsubsection{Lockfile Lifecycle} 

\begin{table}
\small
\caption{Lockfile life cycles across package managers. All behaviors shown here are observed when invoking projects build under the default configuration. Special flags required to achieve the corresponding behavior are given within brackets.}
\label{tab:lockfileLifecycles}
\centering
\begin{tabular}{rrrr}
\toprule
\makecell[r]{Package\\ manager} & \makecell[r]{Lockfile generation} & \makecell[r]{Lockfile-based \\ resolution} & \makecell[r]{Lockfile enforcement \\ on spec file mismatch} \\
\midrule
npm CLI & by default & by default & \makecell[r]{optional \\ (\texttt{npm ci})} \\ 
\hline
pnpm & by default & by default & \makecell[r]{optional \\ (\texttt{--frozen-lockfile})} \\
\hline
Cargo & by default & \makecell[r]{optional \\ (\texttt{--locked}, \texttt{--frozen})} & \makecell[r]{optional \\ (\texttt{--locked}, \texttt{--frozen})} \\
\hline
Poetry & by default & by default & by default \\
\hline
Pipenv & by default & \makecell[r]{optional \\ (\texttt{--deploy}, \texttt{sync})} & \makecell[r]{optional \\ (\texttt{--deploy}, \texttt{sync})} \\
\hline
Gradle & \makecell[r]{optional  \\ (\texttt{--write-locks} \\ with specified lock states)} & by default & by default \\
\hline
Go & by default & by default & by default \\
\bottomrule
\end{tabular}
\end{table}

All \nppm package managers in this study include a function to generate a lockfile, either by default or when explicitly enabled during dependency resolution. At each of these steps of lockfile generation, dependency resolution, and enforcement, different package managers implement different policies, which we detail below. A summary of the key differences are presented in \autoref{tab:lockfileLifecycles}.

\textbf{Lockfile generation.} 
Package managers take different approaches to generate lockfiles. All package managers except for Gradle generate a lockfile by default. For Gradle, a lockfile is generated only when the \texttt{--write-locks flag} is enabled and lock states are explicitly defined in the \manifest.  

\textbf{Lockfile based dependency resolution.} When a lockfile is present at the beginning of a build, it influences the dependency resolution process.
Lockfile based dependency resolution is the property of respecting the locked states when resolving dependencies as long as they do not conflict with the \manifest.
\modified{In npm, pnpm, Poetry, Gradle and Go, the build process respects the lockfile by default and only includes the locked versions. Conversely, with the default build or install command, Cargo and Pipenv resolves dependencies regardless of the locked versions, ignoring the lockfile.}
Afterwards, Cargo and Pipenv silently update the lockfile during the build. To make Cargo respect an existing lockfile, developers have to use \texttt{--locked} or \texttt{--frozen} flags and for Pipenv they can use the \texttt{--deploy} flag or the \texttt{pipenv sync} command.

\textbf{Lockfile enforcement.}
Lockfile enforcement is the property of failing the build if the lockfile conflicts with the \manifest. Here, a conflict means a discrepancy in the checksum or the resolved version.
In the presence of such a conflict, npm CLI and pnpm resolve versions compatible with the \manifest, overriding the lockfile silently (no enforcement).
In contrast, some package managers fail the build in case of a conflict, thus enforcing the lockfile. 
This is the case for Poetry and Gradle where the lockfile is enforced by default, and an error is raised if conflicts are found. 
In npm, only \texttt{npm ci} command enforces the lockfile, and in pnpm, enforcement only happens with the \texttt{--frozen-lockfile} flag. 
In Cargo and Pipenv, the same flags required for lockfile based dependency resolution (\texttt{--locked} and \texttt{--deploy}) also achieve lockfile enforcement.
For Go, lockfile enforcement happens by default because it does not have a separate developer maintained \manifest, and the checksums in go.sum are automatically validated when a package gets downloaded.

\textbf{Summary of lockfile lifecycles}
Except for Gradle, all package managers generate a lockfile by default but vary in how they consider locked versions. Cargo and Pipenv ignore them when resolving dependencies, whereas npm CLI and pnpm use them unless they conflict with the \manifest. Poetry enforces lockfiles strictly, stopping the build if conflicts are detected.

\begin{tcolorbox}[boxrule=1pt,arc=.3em, left=4pt, right=4pt]
  \textbf{Findings from RQ1}: 
  Lockfiles address a fundamental problem of software engineering: knowing and verifying the exact versions of components required to build a software system. Cargo, Go, and npm CLI lockfiles contain all essential dependency information: resolved package versions, package checksums, package sources. However, npm CLI lockfiles are unnecessarily verbose due to excessive metadata. Gradle lockfiles are overly simplistic as they skip mandatory information such as checksums. 
  Simple, streamlined lockfile processes such as Go's are fool proof and recommended as blueprint for the future.
  %Our findings are useful for developers to learn about lockfiles and for package manager maintainers to correctly design and implement effective lockfile features. 
  
\end{tcolorbox}

\subsection{Methodology for RQ2}\label{sec:protocol-rq2}
Next, we aim to analyze lockfile usage across package managers, by selecting a sample of public GitHub projects and collecting the lockfiles within them, if any.

\textbf{Project selection}:
As the study subjects for RQ2, we collect popular public GitHub projects that use one of the \nppm package managers we consider in the study.
\modified{When collecting GitHub projects, we aim to exclude personal toy projects and only target mature, widely-used projects \cite{Kalliamvakou2014perils}.}
To achieve this, we  consider repositories created between September 30, 2019, and September 30, 2024, that have at least 42 stars, 10 contributors, and 300 commits.
To ensure that we only select actively maintained repositories, we further exclude the projects that have not been updated within the last three months.
For npm CLI and pnpm, we select repositories where the primary language is JavaScript or TypeScript.
For Pipenv and Poetry, we include Python projects. For Cargo, Gradle, and Go, we target Rust, Java, and Go projects, respectively.

\textbf{Package manager identification}:
We filter the projects that use each package manager by selecting projects that have the corresponding \manifest in their root file directory.
In JavaScript, there are multiple package managers that can share the same \manifest format and package registry.
For example, the \manifest for both npm CLI and pnpm is \texttt{package.json}, as shown in \autoref{tab:ecosystems}.
For Poetry as well, the \manifest \texttt{pyproject.toml} is compatible with multiple other package managers and packaging tools, such as pip and distutils. However, in Poetry \manifests, we can infer that a project uses Poetry by checking for section names such as \textit{[tool.poetry]}. We exclude projects where \texttt{pyproject.toml} does not contain any Poetry specific sections.
In Go, we consider only projects that use Go modules (introduced in Go 1.11 or later) and identify them by the presence of a \texttt{go.mod} file in the repository. 

\textbf{Lockfile identification}:
Similar to identifying the \manifest, we determine the presence of a lockfile by searching for the relevant lockfile files in the project directory.
If a project has multiple lockfiles, we discard the project to prevent inconsistencies in the results and their interpretation. For example, in the JavaScript ecosystem, if a project contains both \texttt{package-lock.json} and \texttt{pnpm-lock.json}, we exclude it from the evaluation. We further disregard projects with lockfiles from package managers outside the selected \nppm package managers, such as \texttt{yarn.lock} from Yarn.
For Go, we check for the existence of the \texttt{go.sum} file.
\modified{It is our intuition that, for some developers, including a lockfile is related to the maturity of the project. As a project matures, the dependency tree may become more complex, which could lead developers to adopt a lockfile for better dependency management if they have not already done so.}
Therefore, we also collect the project creation date and the earliest lockfile commit date. Using this information, we derive the number of projects that began including lockfiles in version control within the first six months of project creation, as well as the total count of projects that currently have lockfiles.

\textbf{Excluding projects with no dependencies}:
For projects that use a package manager but do not declare any dependencies, lockfiles do not provide any added value. Therefore, we collect the number of dependencies specified in each \manifest and evaluate lockfile usage only for projects that have at least one declared dependency. 
%When determining whether a dependency is declared in the \manifest, we consider all types of dependencies, including optional ones. 
In multi-module projects, if the main \manifest in the root folder does not specify any dependencies, we check the \manifests of sub-modules. If any sub-module declares third-party dependencies, we keep the project.

\subsection{Results for RQ2: \textbf{\rqPatterns}}

\begin{table}
\small
\caption{GitHub projects analyzed to assess the presence of lockfiles in the version control system. The third column shows projects that started including lockfiles in version control within six months of project creation. The final column shows the number of projects that currently commit lockfiles.}
\label{tab:numberOfGitHubProjects}
\centering

%\rowcolors{2}{gray!10}{white}
    \begin{tabular}{rrrr}
    \toprule
    Package manager & \makecell[r]{No of projects \\ with dependency \\ specification files} & \makecell[r]{Projects with lockfiles \\ within 06 months of \\ creation} & \makecell[r]{Projects with lockfiles \\ as of now}\\
    \hline
    npm CLI & \multirow{2}{*}{\npjsts} & \npnpmlockratiobefore\% (\npnpmlockbefore) & \npnpmlockratio\% (\npnpmlock) \\ 
    pnpm & & \nppnpmlockratiobefore\% (\nppnpmlockbefore) & \nppnpmlockratio\% (\nppnpmlock) \\
    Cargo & \npcargo & \npcargolockratiobefore\% (\npcargolockbefore) & \npcargolockratio\% (\npcargolock) \\
    Poetry & \nppoetry & \nppoetrylockratiobefore\% (\nppoetrylockbefore) & \nppoetrylockratio\% (\nppoetrylock) \\
    Pipenv & \nppipenv & \nppipenvlockratiobefore\% (\nppipenvlockbefore) & \nppipenvlockratio\% (\nppipenvlock) \\
    Gradle & \npgradle & \npgradlelockratiobefore\% (\npgradlelockbefore) & \npgradlelockratio\% (\npgradlelock) \\
    Go & \npgo & \npgolockratiobefore\% (\npgolockbefore) & \npgolockratio\% (\npgolock) \\
    \hline
    Total & \nptotalrepos & \nptotallockratiobefore\% (\nptotallockbefore) & \nptotallockratio\% (\nptotallock) \\
    \bottomrule
    \end{tabular}

\end{table}

% description of the table - introduce columns and rows
\autoref{tab:numberOfGitHubProjects} presents the total number of GitHub projects collected for each package manager, as well as the number of projects that publish their lockfile into the version control system. In the second column of the table, we report the number of projects that we collect, per the protocol introduced in \autoref{sec:protocol-rq2}. The third column shows the number of projects that included lockfiles in their GitHub repository, in the early days of the project (within six months after project creation). The last column shows the projects that currently contain a lockfile.

In total, we collect \nptotalrepos projects that include a \manifest from one of the \nppm package managers we consider. 
% We note the specific case of JavaScript repositories, for which we  collect \npjstsall projects that include a \texttt{package.json} file and have at least one dependency, and for which we cannot determine if the \texttt{package.json} is to build with npm CLI or pnpm. 
Within the first six months of project creation, \nptotallockbefore (\nptotallockratiobefore\%) of these projects began committing lockfiles. For example, \npnpmlockratiobefore\% have committed a \texttt{package-lock.json}, and \nppnpmlockratiobefore\% have committed a \texttt{pnpm-lock.json} within six months of the project creation date.
Over time, \nptotallocklater \xspace (\nptotallockratiolater\%) more projects across all package managers have included lockfiles in version control. As of March 28, 2025, \nptotallockratio\% of these projects have committed lockfiles to version control. \modified{This pattern suggests that developers are more likely to commit lockfiles as their projects mature.}

The key finding from \autoref{tab:numberOfGitHubProjects} is that the Go ecosystem has the highest level of lockfile support, with \npgolockratio\% of projects committing the lockfile to the repository.
On the exact opposite, the state of lockfile usage in Gradle is poor with only \npgradlelockratio\% projects committing their lockfile. 
Below, we study these exceptional cases in detail, to check whether they reflect an anomaly in the data or whether they represent intentional design decisions.

% hightlights of the data presented in the table - go gradle, and that there is a trend to increase
For Go, most projects (\npgolockratiobefore\%) begin including \texttt{go.sum} within six months of project creation. After six months only four projects do not commit lockfiles.
Those four projects are:  \href{https://github.com/GoogleCloudPlatform/buildpacks}{\texttt{GoogleCloudPlatform/buildpacks}}, \href{https://github.com/eolinker/apinto}{\texttt{eolinker/apinto}}, \href{https://github.com/go-admin-team/go-admin}{\texttt{go-admin-team/go-admin}}, and \href{https://github.com/shalb/cluster.dev}{\texttt{shalb/cluster.dev}}. These projects have 13, 48, 29, and 35 direct dependencies in their main modules, respectively. The developers of the project, \href{https://github.com/go-admin-team/go-admin}{\texttt{go-admin-team/go-admin}} committed the lockfile at the early stages of the project but later decided to remove it from version control. The reason for removal is not documented in the public repository. The projects \href{https://github.com/GoogleCloudPlatform/buildpacks}{\texttt{GoogleCloudPlatform/buildpacks}} and \href{https://github.com/eolinker/apinto}{\texttt{eolinker/apinto}} have \textit{go.sum} files for some modules but do not contain one for the main module. 

% vendoring
\href{https://github.com/GoogleCloudPlatform/buildpacks}{\texttt{GoogleCloudPlatform/buildpacks}} also uses the Go vendor folder which allows committing dependencies alongside the project, making them accessible at anytime. Vendoring loads dependencies from the local vendor folder instead of resolving them from remote repositories, thus supporting air-gapped builds. Yet, in situations where the local cache cannot be trusted, the go.sum file would still be useful to ensure the integrity of the dependencies.  

\phantomsection
\label{sec:gradle-outlier}
In contrast, only \npgradlelock Gradle projects have a lockfile out of all \npgradle projects, and only one of them (\href{https://github.com/apple/pkl}{apple/pkl}) committed its lockfile within the first six months of project creation. The three Gradle projects that do commit lockfiles, \href{https://github.com/apple/pkl}{\texttt{apple/pkl}}, \href{https://github.com/google/dwh-migration-tools}{\texttt{google/dwh-migration-tools}}, and \href{https://github.com/datastax/fallout}{\texttt{datastax/fallout}}, are all multi-module projects, with 188, 461, and 53 dependencies across their modules, respectively. 
The \href{https://github.com/google/dwh-migration-tools}{\texttt{google/dwh-migration-tools}} project's \href{https://github.com/google/dwh-migration-tools/pull/503}{issue tracker} refers to an internal document that motivates the inclusion of the lockfile on Github.
The project (\href{https://github.com/apple/pkl}{\texttt{apple/pkl}}) also makes use of finer-grained lockfile functionalities, such as \href{https://github.com/apple/pkl/issues/184#issuecomment-1967125262}{selectively updating only certain dependency versions}.

For the five other package managers, npm CLI, pnpm, Cargo, Poetry and Pipenv there is no clear pattern in lockfile usage. Poetry, Pipenv, and Cargo show relatively high adoption rates, with around 50\% of projects committing lockfiles since the early stages and surpassing 70\% as of now.
In the JavaScript ecosystem, lockfiles are widely used but come from multiple package managers. Among them, npm CLI lockfiles appear to be the most common, more so than pnpm lockfiles. However, over time, the number of JavaScript projects committing pnpm lockfiles has doubled, suggesting a growing adoption of pnpm.

\modified{
These observations provide quantitative evidence confirming the findings of RQ1.
First, it confirms that Gradle does not generate a lockfile by default and requires significant effort from developers to set it up, hence the absence of lockfiles. 
Second, Go requires minimal effort to generate and enforce lockfiles, hence the observed prevalence. }  

% summary of possible reasons for the other ecosystems
%Overall, when considering all the package managers, the projects that use a package manager that generates a lockfile by default also include the lockfile in their version control system.  As for the remaining \nptotalnolockratio\%, there could be multiple reasons why developers decided not to commit a lockfile, despite it being automatically generated. These reasons include, the perceived role of lockfiles in different types of projects such as applications or libraries, different dependency management practices or package manager specific guidelines.   

Quantitative data alone does not fully reveal the reasons why developers choose to commit or omit lockfiles. \modified{To complement this initial quantitative analysis across ecosystems, we conduct qualitative interviews with developers to better understand specific cases where lockfiles help them or do not fully satisfy their needs.}

\begin{tcolorbox}[boxrule=1pt,arc=.3em, left=4pt, right=4pt]
  \textbf{Findings from RQ2}:
  Go has an almost 100\% lockfile commit rate, in stark contrast to Gradle, where adoption is close to zero.
  Our empirical results show that generating a lockfile by default increases adoption.  
  We recommend systematically including lockfiles in version control, as they help improve security and contribute to more efficient project builds.
\end{tcolorbox}

\section{Developer Interviews}

Depending on the perceived usefulness and usability of lockfiles, and project specific requirements, developers may choose to actively use lockfiles or totally ignore them. In this section, we qualitatively analyze lockfile usage across projects in each ecosystem to identify differences in how developers perceive and use lockfiles.
We also aim to discover developer expectations as well as the challenges they face when maintaining lockfiles.

\subsection{Research Questions}

%%%%%%%%%%%%%%%%%%%%%%%%%%%%%%%%%%%%%%%%%%%%%%%%%%%

% what advantages
\newcommand\rqBenefits{What benefits do developers perceive in using lockfiles?\xspace}% performance, replicability, security, transparency

\newcommand\rqChallenges{What challenges do developers face when managing lockfiles?\xspace} %versioning, reviewing, maintaining

%%%%%%%%%%%%%%%%%%%%%%%%%%%%%%%%%%%%%%%%%%%%%%%%%%%%%

\begin{enumerate}[label=RQ\arabic*:, ref=RQ\arabic*, start=3]
    
    \item \label{rq:expectations} \textbf{\rqBenefits}
    % the motivations to use lockfiles
    Here, we  study the benefits of lockfiles as perceived by developers and analyze how these perceptions differ across various package managers. Understanding the practical benefits of lockfiles helps clarify their role in secure and reliable software development.
    
    \item \label{rq:challenges}\textbf{\rqChallenges}
    With this RQ, we study the situations where developers struggle to maintain lockfiles, are unsatisfied with existing lockfile features, and find that lockfiles fall short of meeting their expectations. These findings are beneficial for package manager maintainers to improve their lockfile implementations for better usability and effectiveness.
         
\end{enumerate}

\subsection{Methodology for RQ3 and RQ4}

To answer our two research questions on lockfile usage, we select a set of projects from each ecosystem, including some projects that publish their lockfile in the version control system and some projects that do not publish it. We then conduct semi-structured interviews with the top contributor of each project. In the following, we describe the methodology we follow to select the projects, contact developers, interview them, and analyze their responses.

% \begin{figure}
% \centering
% \includegraphics[width=\linewidth]{images/lockfiles_flow.pdf}
% \caption{An overview of the methodology to collect and analyze quantitative and qualitative data}
% \label{fig:figure}
% \end{figure}

\textit{Participant selection}: We aim to conduct interviews with developers who regularly contribute to open source projects, have experience with managing dependencies, and maintain at least one project that uses one of the package managers under study. 

To select projects that publish a lockfile, we proceed as follows.
We first randomly select two projects per ecosystem from the list of projects that publish their lockfile, as identified in RQ2.
%Within each ecosystem, we do not distinguish between different package managers for the same language, such as npm and pnpm for JavaScript, or Poetry and Pipenv for Python.
We do not select Gradle projects that publish a lockfile, since we observe that this is an outlier behavior in the Java ecosystem (cf. \autoref{sec:gradle-outlier}).

For package managers that have a considerable number of projects that do not publish a lockfile, we also select two projects without a lockfile.
Therefore, for npm/pnpm, Cargo, and Poetry/Pipenv, we select a total of four projects each.
We do not select Go projects that do not publish the go.sum, as this is  an outlier behavior in the Go ecosystem.

\modified{For each selected project, we search the public GitHub profile of the top contributor and send an interview invitation if their email address is publicly available. 
As the main contributor of a mature project, we expect these developers to be knowledgeable about lockfiles and be familiar with dependency management best practices.
When developers respond positively to the initial email, we send them the outline of the interview questions in advance so they can confirm their familiarity with the topics.
If the developer's contact information is unavailable, or we do not receive a response within a week, we continue iterating through our randomly selected projects, until we reach our target of two or four participants per category. 
We arrive at this sample size guided by information power \cite{Malterud2016power}.
Information power suggests that the sample size for qualitative interview studies can be assessed based on factors such as the aim of the study, the specificity of the participants, the existing theoretical background, the quality of the dialogue, and the analysis strategy.
As we interview experts on a clearly scoped topic, and the interviewers themselves have background knowledge of the subject, a smaller sample of two to four participants per category provides sufficient information power for our study.}
% This approach allows us to select a diverse group of developers from various backgrounds and location, who have expertise in dependency management and who contribute to projects in various application domains. 

\textit{Participant demographics}: In total, we sent \npinvitations invitations and received 14 responses: two npm/pnpm developers, two Cargo developers, two Poetry/Pipenv and two Go developers who commit lockfiles.
We also received responses from two npm/pnpm developers, two Cargo developers, and two Poetry/Pipenv developers who do not commit lockfiles.
We also interview one developer (\sap) we were acquainted with through personal connections.
In total, we interview \npinterviews developers and \autoref{tab:demographics} presents the demographics of these participants. 
They work in \npdomains different domains and have an average of \npavgexperience years of experience.
For anonymity, we denote each participant with the letter "P" followed by an integer.
% and we do not provide a link to the GitHub project.
% no of dependencies
% The better the coverage of these categories the wider is the range of analyses possible from our data set.

\textit{Interviews}: We conduct 20 minute online interviews with \npinterviews selected developers.
We record the interviews upon consent from the participants.
One interview  (\bevy) was conducted in person at the participant's request, and in this case, only written notes were taken.
The participants were not provided any compensation for their participation.

% questions
We use two different versions of questions for the developers who commit lockfiles and for those who do not.
Both sets of developers receive the same warm-up questions about their software development background, followed by a few interview preparation questions related to basic dependency management practices.
The main interview questions are designed to address RQ3 and RQ4.
Most questions are open-ended, encouraging a natural conversation and giving participants the opportunity to share experiences beyond direct responses.
These questions cover topics such as the reasons for committing or not committing a lockfile, interactions with lockfiles (for developers who do not commit lockfiles, this pertained to analyzing locally created lockfiles), code review workflows, continuous integration and continuous delivery (CI/CD) practices, challenges encountered when using lockfiles, and suggestions for improving dependency management practices.

\textit{Interview response analysis}: 
We use the Whisper speech recognition tool \cite{radford2023whisper} to transcribe \npinterviews interviews with available audio recordings.
If audio recordings are not available, we refer to the written notes. 
\modified{Two authors conduct line-by-line coding of the transcribed interviews or written notes. The same two authors then collaboratively create an affinity diagram, grouping the codes into emerging themes inductively. In cases of disagreement, the two authors consult a third author to help resolve the conflicts.
This process resulted in \npsubthemes sub-themes and \npthemes main themes related to general software development practices, lockfile benefits, developer expectations, and challenges. The final codebook is available at \href{https://github.com/chains-project/lockfiles-comprehensive-study/blob/main/developer_interviews/codebook.xlsx}{https://github.com/chains-project/lockfiles-comprehensive-study}.}
% Content analysis and coding takes place in a spreadsheet, Miro... application.

\begin{table*}
\small
\setlength{\tabcolsep}{6pt} % Optional: Adjust column padding
\caption{Demographics of developers interviewed about their usage of lockfiles. The project domains are selected based on GitHub project descriptions. The experience indicates the number of years of experience of participants in programming in any language.}
\begin{tabular}{p{0.5cm} r r c r rrr r}
    \toprule & 
     & \makecell{Experience \\ (Years)} & 
     \makecell{Lockfile\\Usage} & 
     \multicolumn{3}{c}{Project Stats} & 
     Project Domain \\
    \cmidrule(lr){5-7}
     &            &                             &           & {\footnotesize Stars} & {\footnotesize Commits} & {\footnotesize Contributors} & \\
    \midrule
    \multirow{5}{*}{\rotatebox{90}{npm/pnpm}}%
     & \sap        & ${>}$ 20       & Yes         & 3k     & 110,892       & 320            & {\footnotesize Website development} \\
     & \libre      & 2 - 5      & Yes         & 24.4k  & 2552          & 221            & {\footnotesize Conversational AI} \\
     & \uap        & 2 - 5      & Yes         & 9.6k   & 1240          & 136            & {\footnotesize Network monitoring} \\
     & \grammy     & 5 - 10     & No          & 2.7k   & 817           & 52             & {\footnotesize Bot framework} \\
     & \neocities  & 10 - 20    & No          & 219    & 476           & 21             & {\footnotesize Website deployment} \\
    \midrule
    \multirow{4}{*}{\rotatebox{90}{Cargo}}%
     & \redlib     & 2 - 5      & Yes         & 1.9k   & 1338          & 50             & {\footnotesize Private social media} \\
     & \sniffnet   & 5 - 10     & Yes         & 23.3k  & 2068          & 50             & {\footnotesize Network monitoring} \\
     & \bevy       & 5 - 10     & No          & 38.9k  & 8720          & 1221           & {\footnotesize Game engine} \\
     & \gramrs     & 2 - 5      & No          & 639    & 1087          & 58             & {\footnotesize Bot API library} \\
    \midrule
    \multirow{4}{*}{\rotatebox{90}{\makecell{Poetry\\Pipenv}}}%
     & \wyze       & 5 - 10     & Yes         & 834    & 512           & 27             & {\footnotesize Home Assistants} \\
     & \patito     & 5 - 10     & Yes         & 391    & 413           & 16             & {\footnotesize Data modeling} \\
     & \scrapli    & 10 - 20    & No          & 601    & 500           & 22             & {\footnotesize Network I/O} \\
     & \gym        & 2 - 5      & No          & 708    & 529           & 13             & {\footnotesize Robotics} \\
    \midrule
    \multirow{2}{*}{\rotatebox{90}{Go}}%
     & \chatgpt    & 10 - 20    & Yes         & 670    & 324           & 12             & {\footnotesize CLI for ChatGPT} \\
     & \gomail     & 10 - 20    & Yes         & 948    & 1587          & 19             & {\footnotesize Email library} \\
    \bottomrule
\end{tabular}
\label{tab:demographics}
\end{table*}

\subsection{Results for RQ3: \textbf{\rqBenefits}}
\label{rq3}

One of the main topics we discuss in the interviews with developers who commit lockfiles is their motivation for doing so. From their responses we identify the benefits they expect from using lockfiles. 
Further questions about debugging, code review, and security practices also reveal the direct and indirect benefits lockfiles offer.
We identify benefits across five main areas: build determinism, integrity, transparency, debugging, and security.

\subsubsection{Build determinism benefit} One of the core functionalities of lockfiles is supporting build determinism over time by locking dependency versions.
A majority of  developers who commit lockfiles (\npinterviewreplicratio) mention build determinism as the main reason for committing lockfiles. 
For instance, both Cargo developers who commit lockfiles state that they view the lockfile as a snapshot, and they are confident that the build will be identical every time they use the same lockfile (\textit{``Like I'm 100\% certain that it's going to be identical no matter when I go back to it''}., \redlib).
The Poetry developer, \wyze, mentions the guarantee of build determinism that lockfiles provide across different machines (\textit{``the major benefit I see is that my state is exactly the same when I pull down the code on a new computer. So it's just great for getting everything into the perfect, the same exact state that I was working with before.''})

Resolving dependency versions deterministically is particularly important in CI/CD pipelines to ensure reliable test results across environments and consistent builds before releases.
\sap comments that they enforce and trust the lockfile in their CI/CD workflow (\textit{``When we go through the pipeline in order to test or to run our tests, typically we use npm ci to install the dependencies by trusting the package lock.''})
\wyze mentions that they always make sure every collaborator uses lockfiles in their release pipelines (\textit{``So like in your release pipeline, you want to make sure you're using your lockfile, right? You don't want to just be updating your whole dependency tree. You want to be using your lockfile.''})

\subsubsection{Integrity benefit} All package managers in this study, except Gradle, record dependency checksums in their lockfile.
These checksums allow the package manager to verify, during subsequent installations, that the recorded checksums match with the dependencies being downloaded from the registry.
This verification helps detect potential package tamperings.
\wyze emphasizes its importance, stating, \textit{``If you don't have a hash of your dependency, you don't know what your dependency is. It's just that simple, right? You're flying blind.''}

All developers trust that the checksums recorded in lockfiles are automatically validated by the package manager, and therefore, they do not carry out any additional checks.
\sap note that (\textit{``We trust  the fact that the checksums are being validated, but we'd never really investigate that.''} Similarly, \wyze explained \textit{``My understanding is Poetry already does validation for the hashes. So when I go and I download, when I go and I sync my repository on a new machine, if the hash of the package doesn't match what's in my lockfile, it will throw me a warning.''}.
Poetry keeps a hash of the \manifest content in the lockfile and performs an additional hash validation when checking whether the \manifest has been updated and the lockfile is outdated.
Poetry developers, such as \wyze, appreciate this additional check, 
(``\textit{So it will say your lockfile's out of date. And then it will make me regenerate my lockfile, which I appreciate.}'').

Identifying package tampering through integrity verification requires that releases remain immutable.
In Go, when a package hosted on GitHub is modified after publication, such as through re-tagging or force-pushing, it can lead to checksum failures, as the version stored in the Go module proxy may differ from the modified release.
A Go developer described this as a trade-off between the convenience of releasing and the security.
(\textit{``it's good on the one hand that you will always, like from a security perspective, make sure that you will always get the code that is basically stored on the proxy. And if you don't have that, something is wrong and the checksum will give you an error. But then again, it can cause issues if you like want to re-release the same version. And yeah, it's like convenience versus security''}, \gomail)

\subsubsection{Transparency benefit} Lockfiles offer transparency into all dependencies resolved in a project, including indirect dependencies which are not visible in a developer maintained \manifest.
When developers commit lockfiles, they can inspect changes to the dependency tree during code reviews and detect any unexpected or potentially malicious changes. 

Both Cargo developers who commit lockfiles mention that they prefer to verify pull requests that modify the dependency tree by re-generating the lockfile locally to ensure the changes are reproducible.
(\textit{``Since I prefer to double check, I clone their fork locally, and I run cargo update, so that I'm sure that the lockfile is reflected on the actual changes, and, you know, maybe they had something malicious, so it's better to check.''}, \sniffnet.) \redlib confirms it stating \textit{``Typically what I will do is, I'll pull their change into a branch. I will recreate the change that they did to the cargo.toml and then see if my lockfile changed in the exact same way.''}

Lockfiles, by exposing indirect dependencies, help developers monitor and manage dependency growth over time.
\gomail recall a time where they advised a collaborator to remove an unnecessary dependency \textit{``So I said, okay, let's not just import this whole dependency just for one Go file. Let's just import the code over because it's probably not changed anyways. And yeah, so usually I check the dependencies and basically question is this dependency needed or not''}.

\subsubsection{Debugging benefit} Almost all developers have experienced breaking dependency updates and semantic versioning (semver) violations in their projects.
Developers who commit lockfiles use them to help in debugging breaking changes.
\libre recollect how they copied a previous version of a lockfile to resolve a breaking update, \textit{``That was one time I literally manually copy-pasted. I went exactly to where that package was resolved before, and it was working, and I literally replaced it''}. 

Another advantage of using locked versions is that developers can explicitly control when to allow dependency updates, preventing unexpected breakages.
\patito mention that they update the lockfile only when they are prepared to handle potential breaking changes.
(\textit{``From time to time, I run poetry updates with the intention of getting the latest and being in a mindset where I now have the surplus time and energy to handle breaking changes or issues that could arise from using the latest versions.''})
\sniffnet mention how they have to enforce the lockfile to prevent a dependency from getting updated because it does not follow the semantic versioning guidelines.
(\textit{``So basically, it was the same problem. Semantic versioning was not respected by them. So I updated the README, and saying that if they download from cargo, they should use cargo install -{}-locked. Because in that way, I'm sure that they're using the dependency in the lockfile.''}). 

In Cargo, there is the yanked release issue, when dependency is no longer available by version, but remain available by checksum.
The Cargo dependency resolver ignores yanked versions unless they are specified in the lockfile, occasionally causing the build to fail.
\redlib discuss how they overcome the yanked version problem in Cargo through lockfiles, \textit{``If you have the lockfile, you can always build it, even if it's yanked. But if you don't have a lockfile and it gets yanked, you cannot under any circumstances build it''}.

\subsubsection{Security benefit}
Another key point raised by the developers is that lockfiles are indirectly used by other security scanning tools.
For instance, \sap mention using the \href{https://www.blackduck.com/}{Black Duck} software component analysis tool, which relies on lockfiles (\textit{``Black Duck uses a list of vulnerabilities based on the package names, on their versions, and reports on a regular basis on our products to know if one of the dependencies of this product is having a known vulnerability. So that's one thing which leverages the package-lock.json''}). When a lockfile is present, dependency management bots such as Dependabot can also detect and flag vulnerabilities within indirect dependencies (\textit{``It's definitely beneficial to have a lockfile for the purposes of Dependabot. Yeah. Dependabot is what I rely on heavily for making sure I'm dealing with supply chain security properly... So Dependabot will scan the lockfile in order to determine the whole dependency tree, and then it wll warn you for CVEs even within transitive dependencies''}, \wyze).  

% Having the lockfile in the version control system is also seen as an advantage, as it can serve as proof of changes. "\textit{I like having that history just for proving, oh, hey, I patched this bug in this dependency. Here's the commit. Here's the lockfile.}" (\wyze), and "\textit{I've had situations where I could look back in my Git history and see, okay, well, this package was working}" (\libre).

\begin{tcolorbox}[boxrule=1pt,arc=.3em, left=4pt, right=4pt]
  \textbf{Findings from RQ3}:
   Lockfiles offer benefits across various dimensions of software development, including build determinism, integrity, transparency, debugging, and security. Developers who commit lockfiles, regardless of the package manager, are generally aware of these benefits. In each ecosystem, many developers have experienced unique software development problems where lockfiles is the solution. Overall, developers who use lockfiles see them as an essential component for effective dependency management.
\end{tcolorbox}

\subsection{Results for RQ4: \textbf{\rqChallenges}}
\label{rq4}

As part of our interview protocol, we ask developers who do not commit lockfiles about their reasons, and those who do commit lockfiles about the challenges they face when using them. From their responses, we identify several common challenges, including indirect locking with libraries, managing dependency updates, limited readability, cache invalidation, and the learning curve associated with lockfile practices.

\subsubsection{\modified{The challenge of locking library dependencies}} 

The main reason mentioned by four out of six developers who do not commit lockfiles is that their project is a library rather than an application or executable. These projects are in Python, Javascript or Rust and are in various application domains. The challenge they face is that even if they lock the versions of dependencies in their project, they cannot publish a lockfile to the package registry, and downstream dependents could not enforce the lockfile for their library. As a result, they choose not to enforce a lockfile in their development environment, as doing so would give them a false sense of security. 
(\textit{``So basically it is better not to commit a lockfile because then you kind of run into these issues faster. And, uh, yeah, you don't have this false sense of security where you think it's all locked, but in reality, everybody who uses it doesn't lock it.''}, \grammy). Following the same reasoning, \bevy explain that they prioritize the experience of the users of their library over their own developer experience, and as a result, they do not commit lockfiles.

Similarly, some developers view determinism as a concern of downstream users who depend on their library.
(\textit{``You could make your end thing deterministic, but as a library author, for me, it's, that's not really my problem.''} \scrapli). 

\gramrs do not currently commit lockfiles and they mention that they would reconsider their decision based on the updated Cargo guidelines introduced in 2023. They refer to the  \href{https://stackoverflow.com/questions/62861623/should-cargo-lock-be-committed-when-the-crate-is-both-a-rust-library-and-an-exec}{StackOverflow thread} \textit{Should Cargo.lock be committed when the crate is both a Rust library and an executable}. (\textit{``This was the previous recommendation on StackOverflow, but has since changed. I have not revisited this choice, but if I were to start a new project today, I would follow the latest recommendations and defaults''}, \gramrs.)

\subsubsection{Slowing down dependency updates}

A second challenge developers face with lockfiles is that version locking can slow down frequent dependency updates. Almost all developers use bots such as Renovate and Dependabot to manage dependency updates, which also handle automatic lockfile updates. However, they are concerned about the lag associated with updating dependencies through bot-generated pull requests. P8, in particular, mention that they prioritize frequent dependency updates over deterministic builds to provide a better experience for their library users.

\patito mention that, with Poetry, if a lockfile references a version of a dependency that has been yanked, it results in an error, unlike Cargo. However, if they do not have a lockfile, Poetry will resolve a newer version, without raising an error. They explain the issue with an example.
\textit{``Now, if you, at 9 a.m. in the morning, let's say a library pushes a new version. At 9.30, you create a lockfile with that version locked. And at 10 a.m., they delete that version from PyPI. Well, then anybody who syncs using that lockfile hits a big issue. Because that version is no longer available''}. At the same time, they acknowledge that this behavior can sometimes be helpful. \textit{``it reports it as yanked, and therefore doesn't install it. So you get an early error, which in some situations is desirable''}.

In npm, the current state of project dependencies is referred to as the ideal state. For projects without a lockfile, npm CLI always resolves dependencies to this ideal state at the time of installation. \neocities prefer resolving this ideal state rather than relying on a potentially outdated lockfile.
\textit{``Because you're always resolving the ideal tree, like at a given moment, that tends to be a better resolution than whatever state your lockfile is accumulated over six months or whatever''}. They also believe that continuously resolving the ideal tree helps fixing bugs and vulnerabilities more effectively. \textit{``Oftentimes, either you'll just miss the defect because it will be introduced for a short period of time and then got published with a fix. And so you just won't even notice it just because you're always resolving the ideal tree''}. 

\subsubsection{Human readability challenge}

All five npm/pnpm developers mention that they find the lockfiles too long, verbose and hard to read. For example, \sap state \textit{``it's quite wild what you can find inside the package lock. So it's not really readable''}, whereas \libre mention how this verbosity can complicate the code reviews and continuous integration, \textit{``Git is kind of overloading and taking up a bunch of memory. And is even having trouble showing them all''}. \neocities also comment on the human readability of lockfiles, \textit{``save us from the insanity that these machine generated lockfiles, which are just, you know, they're human readable, but they're not because they're so big''}. 
Poetry/Pipenv developers raise the same issue where \wyze explain,
\textit{``it's like a huge JSON file. So it's kind of annoying to review the diff. And I don't get a lot of value out of reviewing the diff, necessarily.''}. \patito add that they also find it difficult to code review the lockfile, \textit{``I will be much more likely to review any changes to a pyproject.toml or, you know, direct dependency file. So I'd say it's to some degree a weakness of lockfiles that they are so big''}.

On the other hand, none of the Cargo or Go developers raise issues related to the readability. Go developer \gomail specifically mention they find reviewing go.mod files very easy. 
\textit{``But usually my Go mod files are very like, they do not go over one page of my screen''}.
\scrapli who develop in  Python and do not use lockfiles also comment that they get a better developer experience with Go, \textit{``It's basically almost the same because go.mod is also kind of very minimalistic, So I guess you get all of the benefits of like, you know, pip lockfile or poetry lockfile or whatever with go without having to think about it.''} This confirms the fact that, depending on the lockfile implementation details, such as the content included in lockfiles, developers either effectively use them for manual code reviews or tend to overlook them entirely.

When asked about the information they consider during code reviews of lockfiles, the majority of developers who commit lockfiles mention that they focus on the name, URL, version of the package, and the number of indirect dependencies it includes. \modified{The developers do not pay attention to the exact details of indirect dependencies, and trust the package manager to handle them (\textit{``I'm peripherally aware that the lockfile has changed. You could say it's on my mind, but at some point you just have to trust the process, there's too many dependencies for me to go through everything, for me to do code review on everything.''}, \patito).} A consistent observation we make is that none of the developers pay attention to the checksum of the dependencies in the lockfile. (\textit{``I've never actually checked checksums''}, \sniffnet). \modified{They do not attempt to read or manually verify the checksums. Instead, they rely on the package manager to automatically verify them.} Following the same reasoning, Go developers recommend the Go's approach of having checksums in a separate file.
(\textit{``To be honest, I really don't care where the checksum is stored. It makes it clean to have it separated because I only have to put it into the Go mod file and then the checksum will be generated automatically in the sum file. So that's like having a look at your Go mod file is much cleaner.''}, \gomail).

\subsubsection{Cache challenge}
A common issue faced by all developers is cache invalidation. \bevy remark that \textit{``Cargo caching mechanisms are buggy''}, whereas \sap mention that they have to recreate the lockfile to address cache invalidation problems, \textit{``I would say most of the time what people do is they delete the node modules folder, they delete the package lock, and then they retry and install''}. Following up on cache invalidation, \libre mention that they occasionally run into operating system specific issues which cannot be solved by simply deleting the cache (\textit{``Just working across different operating systems, um, you know, also on macOS. I've had issues where deleting, like doing that same process, deleting the node modules folder, created issues with npm and then nothing would install.}''). Go developers also face cache related issues with Go proxy when packages get re-released under the same version (\textit{``Go proxy already had that basically broken release downloaded and stored in its proxy cache. So people were like opening issues that they had checksum issues''}, \gomail).

\subsubsection{Learning challenge}
Another point raised by developers is the learning curve of lockfile related practices across different package managers. \chatgpt compare Gradle and Go noting that they find it easier to learn and understand dependency resolution practices in Go, \textit{``it's very, very simple. It doesn't have much of a learning curve, such as like Maven or, Gradle''}. \scrapli also comment on this \textit{``I think Go is just an easier, it's just easier. It works better. And, and, you know, it's a newer language and lessons learned''}. 

npm developers often mention that the logic behind dependency resolution and lockfile creation processes are unclear (\textit{``maybe an understanding of how it constructs the lockfile. Cause it's still kind of a mystery to me''}, \libre). Many also comment on the unhelpfulness of error messages when debugging lockfile related issues (\textit{``The error message that you get is really cryptic.''}, \sap). They note that resolving these issues requires an in-depth understanding of the underlying processes, and the debugging itself is extremely time consuming (\textit{``Some of these issues I've had with it are just so hard to debug until I do some very thorough digging of what's going on''}, \libre). \gomail compare npm CLI with Go and explain that the minimalism makes the dependency resolution process in Go more transparent and easier to understand. 
\textit{``You just say npm install and like a wall of text goes down and you have a gazillion packages installed and you don't really know what happens and why those are installed. Compare that to Go, which is, in my opinion, very open and very clear on what dependencies you are using''.} 

\begin{tcolorbox}[boxrule=1pt,arc=.3em, left=4pt, right=4pt]
  \textbf{Findings from RQ4}: Some developers find it challenging to manage lockfiles:  readability challenges due to the verbosity of lockfiles, delays in dependency updates, or the challenge of enforcing the lockfile for a library. Overall, developers agree on the fact that lockfile challenges are better managed in Go than in JavaScript or Python. 
\end{tcolorbox}

\section{Implications of Results}

In our study, we have analyzed package managers and lockfiles from diverse perspectives, reading code and documentation, comparing features and contents, and interviewing users.
This gives us a unique perspective for reflection on the principled engineering of lockfiles for a healthy software supply chain.

\subsection{Recommendations for package manager users}

For developers who use package managers, we provide two main recommendations based on our analysis: selecting a package manager with a suitable lockfile feature, and following best practices for lockfile usage to get the most out of them.

\textbf{[Recommendation \#1] We recommend selecting package managers with human readable lockfiles that are enforced by default.}

In the JavaScript ecosystem, we study two package managers: npm and pnpm. npm includes extensive metadata for each dependency, making its lockfile more verbose compared to pnpm. We observe that many developers struggle with maintaining and reviewing npm lockfiles.  Therefore, we recommend pnpm for JavaScript developers who prefer manually monitoring lockfile changes. 
% On the other hand, pnpm does not include the source link for downloaded packages, which is a significant limitation for validating dependency provenance. 

The main difference between Poetry and Pipenv is in the lockfile-based resolution and enforcement. Poetry enforces the lockfile by default with the install action and throws an error if the lockfile needs to be updated whereas Pipenv silently updates the lockfile when a newer dependency version within the specified range in the \manifest is available.
Therefore, we recommend Poetry over Pipenv for Python projects operating in untrusted environments, such as those relying on untrusted local caches, and requiring frequent dependency integrity checks.

\textbf{[Recommendation \#2] We recommend committing lockfiles for all projects.}

Library developers face the same security risks and collaborative development challenges as application developers. As explained by library developers in \autoref{rq4}, the perceived cost of committing a lockfile is a false sense of security: they do not encounter in-range breakages themselves, while the library users still may.
To mitigate such unexpected issues, library developers can run the library’s test suite both with and without enforcing locked versions in their CI/CD pipelines to detect any in-range breakages. Such an approach may allow them to commit a lockfile while still ensuring that the library remains free of in-range breakages.
Another observation we made is that library developers often skip committing lockfiles because they believe determinism is not a primary concern for them. However, as discussed in \autoref{rq3}, lockfiles offer several other practical benefits such as enabling distributed integrity checks, which library developers can also take advantage of. Therefore, we recommend committing lockfiles for all projects.

\subsection{Recommendations for package manager developers}

Drawing inspiration from the best lockfile features, package manager developers can refine the lockfile functionality in existing package managers to improve the overall developer experience. Moreover, new package managers continue to emerge, introducing innovative features and addressing limitations in traditional ones. These new tools can adopt best practices around lockfiles and have the potential to influence a broad developer community. Based on our analysis, we suggest three recommendations to guide the future development of package managers.

\textbf{[Recommendation \#3] We recommend future package managers to prioritize developer experience.}

Having multiple configuration options for lockfiles gives developers more flexibility and control.
Yet, it can also discourage them from using the basic functionalities of lockfiles due to the added complexity.
For less experienced developers, setting up appropriate configuration can be particularly confusing.
For instance, in Gradle, developers have to enable a flag, update the \manifest with lock states, and select an appropriate mode to generate a lockfile.
On the other hand, sensible defaults and strict enforcement simplifies the process of using lockfiles and makes it easier to follow best practices.
In Go, the complexity of lockfiles is remarkably reduced to a bare minimal developer intervention.

\textbf{[Recommendation \#4] We recommend lockfiles to include only essential content.}

When designing lockfiles, package managers should avoid duplicating information already present in the \manifest. They should also exclude data that is irrelevant to dependency locking.
In npm, for example, including data such as licenses and funding information in the lockfile adds little to no perceived value while unnecessarily increasing  code reviewing cost. 
At the other extreme, Gradle presents a serious lockfile anti-pattern by omitting essential details such as checksums.
We recommend lockfiles to include the following mandatory details: resolved package versions, checksums, URLs for downloaded packages, and a clear distinction between direct and indirect dependencies.
Any additional metadata should be kept as brief as possible, and ideally included only for reproducibility purposes, such as OS, engine, and language versions.
We also encourage separating checksums into a separate file, similar to the approach of Go, to improve the human readability of lockfiles. 

\textbf{[Recommendation \#5] We recommend package managers to generate lockfiles by default.}

Lockfile generation should be enabled by default as lockfiles provide unique software engineering benefits, including integrity verification, reproducibility, and performance improvements, all no cost if enabled by default.
We claim that dependency resolution should respect the locked versions by default and should never silently update the lockfile to prevent any unexpected build changes. This is essential to detect resolution based software supply chain attacks.
As example to follow, Poetry respects locked versions and warns when the lockfile is out of date without silently updating it, we observed that developers who care about project security do appreciate this behavior. On the other end, when the lockfile is not generated by default, as with Gradle, developers seldom use them. 

\section{Threats to Validity}

% When selecting the projects, we only considered those that are actively maintained and have been updated within the last three months. Therefore, we expected them to use a recent version of the package manager. However, developers might be using different versions of the package manager in their personal setups. Since it is not possible to directly retrieve this information from the project settings, verifying the exact versions is not straightforward. This introduces the possibility that subtle changes to the lockfile functionality between different package manager versions may not have been considered in this study.

\modified{One threat to validity in our study is the selection of keywords used to analyze the implementation of package managers, which may not be exhaustive. We mitigated this threat by combining multiple sources, including both source code and documentation.}

\modified{
% The list of projects selected for the quantitative analysis is neither exhaustive nor balanced. 
The imbalance in the number of projects that use a particular package manager is another potential threat to validity. To minimize bias, we applied the same selection criteria across all package managers and included all projects that satisfied these criteria. 
% However, on the day of data collection, the number of Pipenv projects was relatively low compared to the number of projects collected from other package managers, due to its limited adoption. 
}

\modified{We acknowledge that our dataset and the developers who participated in the qualitative analysis may not represent the entire software engineering practices or the full spectrum of developer perspectives. We also did not interview Gradle developers or Go developers who do not commit the go.sum file, as we considered them outside the norm. This is a threat to validity, as the perspectives of these outliers were not captured. Nevertheless, our results, collected from developers across different ecosystems, remain valid regarding the standard practices within each ecosystem. They provide a broad overview of lockfiles as a general concept across package managers.}

\section{Related work}

To the best of our knowledge, our work is the first systematic analysis of lockfiles in the literature.
Some studies occasionally mention them in the broader context of dependency management.

\subsection{Dependency resolution and package managers}
%functionality: technology for dep management

One of the key responsibilities of a package manager is resolving compatible versions for all the dependencies in a project. Zacchiroli \etal \cite{abate2020dependency} review the usage of dependency resolving algorithms, including specialized constraint solving algorithms as well as ad-hoc dependency tree based algorithms. They conduct a census of dependency resolution mechanisms across 18 package managers, using version locking as one of the criteria for comparison.
Pinckney \etal \cite{pinckney2023maxsmt} introduce a unified framework built on top of Max-SMT solvers to resolve dependencies more systematically, moving beyond ad-hoc algorithms.

% dependency conflicts
Several studies focus on dependency conflict issues in different ecosystems.
For example, \cite{wang2020watchman} identifies factors that lead to conflicts, whereas \cite{wang2022smartpip} highlights issues related to inefficiency and excessive resource  usage by the dependency resolution strategies in Python.
\cite{wang2019stacktrace} introduces a tool to generate crash stack traces for dependency conflicts in Java.
Patra \etal \cite{Patra2018conflictjs} analyze dependency conflicts in JavaScript that arise from namespace collisions. \cite{wang2025peer} presents an in-depth study of conflicts between peer dependencies in npm. Similarly, in the Go ecosystem, Wang \etal \cite{wang2021gopath} study conflicts caused by the coexistence of two library referencing modes: Go PATHs and Go modules.

\modified{Decan \etal \cite{decan2018npm} analyze vulnerabilities in the npm dependency network, focusing on how they are discovered and addressed.} On the same lines, Liu \etal \cite{liu2022demystifynpm} analyze vulnerability propagation within dependency trees by applying npm-specific dependency resolution rules, rather than relying on simple reachability analysis for indirect dependencies. They recommend managing dependencies with lockfiles.
He \etal \cite{futile2025he} study the effect of dependency pinning in npm projects. They find that local pinning leads to more security vulnerabilities due to bloated and outdated dependencies. 
They also suggest that the risk of malicious package updates can be reduced when a small set of core dependencies pin their versions and keep them updated regularly. Rahman \etal \cite{rahman2025pinning} confirm the results of He \etal \cite{futile2025he}, and recommend lockfiles for effective dependency management. 

These previous studies analyze dependency resolution mechanisms either directly or in the context of vulnerability detection. Our work complements these works with an analysis focused on  the impact of locking of dependency resolution across package managers.

\subsection{Dependency management practices in open source software}

% breaking changes
Several empirical studies provide insights into the dependency management practices of open source software projects. 
Bogart \etal \cite{Bogart2021} discuss practices related to breaking changes across different package managers, and highlight the use of lockfiles to lock versions of indirect dependencies as a means to stabilize projects.
Breaking updates have been well studied in the literature \cite{ochoa2021bad, frank2924bump, frank2024bg}. 
Venturini \etal \cite{venturini2023dependubreaknpm} quantitatively evaluate the impact of breaking updates on dependent packages in the npm ecosystem.
They also note the advantages of using lockfiles to track all resolved packages and avoid in-range breaking changes. 

% Rust
Schueller \etal \cite{Schueller2022rust} curate data from the Rust ecosystem over eight years, capturing developer activity, library dependencies, and usage trends.
Another empirical analysis in Rust \cite{Hao23} reveals that 46\% of packages adopted yanked releases and the proportion of yanked releases keeps increasing over the years. 
In Cargo, such yanked releases can be resolved only if a lockfile is present.

% security
There is a body of work that analyzes security practices and threats associated with dependency management.
Gu \etal \cite{gu2023investigating} conduct a systematic analysis of security threats in package registries; Maven, PyPI, npm, Cargo, NuGet, and Go. 
Rahman \etal \cite{rahman2025vulnerabilitydataproblempredicting} propose a metric to evaluate vulnerabilities in dependencies based on the time it takes for a fix to be applied.
Bos \cite{bos2023review} reviews supply chain attacks and discusses a type of attack that originates from lockfile tamperings.
One such tampering is adding a duplicate entry to the lockfile with a modified package and its new checksum.
As most lockfiles are tediously long for thorough review, such tampering goes undetected leading to supply chain attacks. 
Vaidya \etal \cite{vaidya2019security} characterize attacks in npm and PyPI ecosystems.
Their findings suggest that with proper tooling and practices it is possible to mitigate these attacks to a great extent.
The authors discuss the npm CLI lockfile \texttt{npm-shrinkwrap} that allows indirect dependency locking.
They mention that \texttt{npm-shrinkwrap} can prevent future attacks from new packages through dependency locking.
Kabir \etal \cite{kabir2022securitybest} study security best practices in software development and consider committing lockfiles as one of the best practices. 
Their study focuses exclusively on the npm ecosystem, whereas our work spans multiple ecosystems and centers specifically on lockfiles. 

Most of the related work on dependency management practices discusses lockfiles only briefly.
Our contribution is unique by being entirely focused on lockfiles, and providing novel software engineering knowledge on their usage and adoption.

\subsection{Adoption of dependency management tools within development teams}
% how people use them: devs interviews about dep management

% signing 
Previous works have also used combined qualitative and quantitative methods to learn about developers' adoption of dependency management tools.
Schorlemmer \etal \cite{schorlemmer2024signing} study the adoption of software signing in four package registries; Maven, PyPI, DockerHub and Huggingface.
Kalu \etal \cite{kalu2024industry} conduct a follow-up interview study to understand software signing in practice.
They identify that strict signature rules increase the quantity of signatures, and that registry policies have an impact on the developer decision to adopt proper signing.  
% GitHub
% Saroar \etal \cite{saroar2023githubactions} survey 90 developers to understand how they use, choose, and troubleshoot GitHub Actions.
% Holtgrave \etal \cite{holtgrave2025attributing} analyze different attack types that can be performed by impersonating GitHub user accounts.
Bifolco \etal \cite{bifolco2024githubdep} assess the accuracy of the GitHub dependency graph in Java and Python projects. They use lockfiles as one source of truth to assess the GitHub dependency graph. 

%bots
Dependency management bots such as dependabot, renovate and Greenkeeper automate the process of updating \manifests and lockfiles when new dependency versions are available. 
Alfadel \etal \cite{alfadel2021dependabot} and Mohayeji \etal \cite{mohayeji2023dependabot} evaluate how developers respond to security updates suggested by Dependabot. \modified{Mohayeji \etal \cite{Mohayeji2025dependabot} also investigate how Dependabot can help mitigate vulnerabilities, noting that it uses lockfiles to create a comprehensive graph of dependencies.}
Rombaut \etal \cite{Rombaut2023} study whether Greenkeeper reduces developer effort or introduces unnecessary workload.
They mention lockfiles as a way to overcome in-range breaking changes, and also briefly mention about the challenge library developers face for sharing lockfiles.
Our qualitative analysis confirms that developers do use bots to maintain \manifests and lockfiles up to date.

Our novel findings on lockfiles, their contents and their adoption in practice contributes to the broader body of knowledge about the state of practice regarding the state of the art of dependency management tools.

\section{Conclusion}

We have conducted the first in-depth study  of lockfiles across \nppm package managers. We analyzed their documentation and implementation, and collected quantitative data from public GitHub projects and qualitative data from developer interviews. 
We systematically documented how different package managers follow different approaches to generate and enforce lockfiles, and these decisions ultimately result in varied levels of developer satisfaction.
We further identified the reasons why developers find lockfiles useful: to achieve deterministic builds, to perform integrity checks, to simplify debugging. We also unveiled the unique challenges developers face when using lockfiles: readability issues, indirect dependency locking limitations, and delayed dependency updates. 
Based on these observations, we provided \nbrecommendationsus recommendations for package manager users and \nbrecommendationsdev recommendations for package manager developers.

Today, only the Go ecosystem systematically generates and adds lockfiles in version control system. The Go lockfile is particularly readable, concise and clearly separates the list of dependencies from their checksums, making them easier to understand for developers. On the other hand, lockfiles are almost absent from the Java ecosystem. Out of the \nppm package managers we study, only Gradle does not generate a lockfile by default. Meanwhile,  Maven, the other major package manager for Java does not have a lockfile at all. We recommend the Maven community to add this feature and learn from the best practices to design an informative and usable lockfile.

% As future work, one important direction is to study the adoption of lockfiles within the software supply chain of machine learning models \cite{Wang2024}, and discover ways to improve the readability of lockfiles by reducing the number of dependencies, e.g. through code transplantation \cite{Jahanshahi2025}.
As future work, we shall experiment with ecosystem locking to capture the overall state of the package registry, instead of locking individual packages. 
We shall design and implement support for developers to toggle indirect lockfile enforcement and to allow library authors to publish a lockfile to the package registry, for giving downstream users the flexibility to choose between locking dependencies or allowing floating versions. 
Clearly, \href{https://www.imdb.com/title/tt1199099/characters/nm0000457}{\color{black}the destiny} of software integrity lies on lockfiles.

\section*{Acknowledgments}

We thank all the interview participants for dedicating their time for this work. We acknowledge the valuable insights provided by Ruy Adorno, a software engineer at \href{https://www.vlt.sh/company}{vlt}. We also thank Ian Arawjo for feedback on the interview protocol and early versions of this paper.

\section*{Declarations}

\subsection*{Funding}
This work is supported by the CHAINS project funded by Swedish Foundation for Strategic Research (SSF), by the Wallenberg Autonomous Systems and Software Program (WASP) funded by the Knut and Alice Wallenberg Foundation, and by IVADO and the Canada First Research Excellence Fund.

% \subsection*{Author Contributions}

% \textbf{Yogya Gamage} contributed to the experimental design, data collection, data analysis and  writing. 
% \textbf{Deepika Tiwari} contributed to data analysis and writing. \textbf{Martin Monperrus} contributed to data analysis and writing. \textbf{Benoit Baudry} contributed to the experimental design, data analysis and  writing.

\subsection*{Data Availability Statement}

All materials related to the qualitative analysis that cannot be traced back to the interview participants, including interview invitation emails, interview protocol, the codebook,  as well as all data from the quantitative analysis, are available at https://github.com/chains-project/lockfiles-comprehensive-study.

% \subsection*{Generative AI}

% \modified{Generative AI was not used for the generation of any part of the content in this paper or for data analysis. Grammarly, a tool that uses AI, was used for spell checking, grammar correction, and improving writing clarity.}

\subsection*{Ethical Considerations}

\modified{This paper includes minimal-risk data collected from developers during interviews. Participation in the study was entirely voluntary. All data collected from participants were anonymized and were handled in accordance with the \href{https://www.priv.gc.ca/en/privacy-topics/privacy-laws-in-canada/the-privacy-act/}{Privacy Act}. Prior to the study, we obtained approval from our organization’s Research Ethics Board (REB).
% Authors who interacted with the interviewees completed the training procedures on research ethics.
}

% \subsection*{Conflict of Interest}
% None
% \subsection*{Clinical Trial Number}
% Not applicable
% \bibliographystyle{unsrtalpha}
\bibliography{biblio}


\begin{thebibliography}{48}
\ifx \bisbn   \undefined \def \bisbn  #1{ISBN #1}\fi
\ifx \binits  \undefined \def \binits#1{#1}\fi
\ifx \bauthor  \undefined \def \bauthor#1{#1}\fi
\ifx \batitle  \undefined \def \batitle#1{#1}\fi
\ifx \bjtitle  \undefined \def \bjtitle#1{#1}\fi
\ifx \bvolume  \undefined \def \bvolume#1{\textbf{#1}}\fi
\ifx \byear  \undefined \def \byear#1{#1}\fi
\ifx \bissue  \undefined \def \bissue#1{#1}\fi
\ifx \bfpage  \undefined \def \bfpage#1{#1}\fi
\ifx \blpage  \undefined \def \blpage #1{#1}\fi
\ifx \burl  \undefined \def \burl#1{\textsf{#1}}\fi
\ifx \doiurl  \undefined \def \doiurl#1{\url{https://doi.org/#1}}\fi
\ifx \betal  \undefined \def \betal{\textit{et al.}}\fi
\ifx \binstitute  \undefined \def \binstitute#1{#1}\fi
\ifx \binstitutionaled  \undefined \def \binstitutionaled#1{#1}\fi
\ifx \bctitle  \undefined \def \bctitle#1{#1}\fi
\ifx \beditor  \undefined \def \beditor#1{#1}\fi
\ifx \bpublisher  \undefined \def \bpublisher#1{#1}\fi
\ifx \bbtitle  \undefined \def \bbtitle#1{#1}\fi
\ifx \bedition  \undefined \def \bedition#1{#1}\fi
\ifx \bseriesno  \undefined \def \bseriesno#1{#1}\fi
\ifx \blocation  \undefined \def \blocation#1{#1}\fi
\ifx \bsertitle  \undefined \def \bsertitle#1{#1}\fi
\ifx \bsnm \undefined \def \bsnm#1{#1}\fi
\ifx \bsuffix \undefined \def \bsuffix#1{#1}\fi
\ifx \bparticle \undefined \def \bparticle#1{#1}\fi
\ifx \barticle \undefined \def \barticle#1{#1}\fi
\bibcommenthead
\ifx \bconfdate \undefined \def \bconfdate #1{#1}\fi
\ifx \botherref \undefined \def \botherref #1{#1}\fi
\ifx \url \undefined \def \url#1{\textsf{#1}}\fi
\ifx \bchapter \undefined \def \bchapter#1{#1}\fi
\ifx \bbook \undefined \def \bbook#1{#1}\fi
\ifx \bcomment \undefined \def \bcomment#1{#1}\fi
\ifx \oauthor \undefined \def \oauthor#1{#1}\fi
\ifx \citeauthoryear \undefined \def \citeauthoryear#1{#1}\fi
\ifx \endbibitem  \undefined \def \endbibitem {}\fi
\ifx \bconflocation  \undefined \def \bconflocation#1{#1}\fi
\ifx \arxivurl  \undefined \def \arxivurl#1{\textsf{#1}}\fi
\csname PreBibitemsHook\endcsname

%%% 1
\bibitem[\protect\citeauthoryear{Alfadel et~al.}{2021}]{alfadel2021dependabot}
\begin{bchapter}
\bauthor{\bsnm{Alfadel}, \binits{M.}},
\bauthor{\bsnm{Costa}, \binits{D.E.}},
\bauthor{\bsnm{Shihab}, \binits{E.}},
\bauthor{\bsnm{Mkhallalati}, \binits{M.}}:
\bctitle{On the use of dependabot security pull requests}.
In: \bbtitle{2021 IEEE/ACM 18th International Conference on Mining Software Repositories (MSR)},
pp. \bfpage{254}--\blpage{265}
(\byear{2021}).
\doiurl{10.1109/MSR52588.2021.00037}
\end{bchapter}
\endbibitem

%%% 2
\bibitem[\protect\citeauthoryear{Abate et~al.}{2020}]{abate2020dependency}
\begin{bchapter}
\bauthor{\bsnm{Abate}, \binits{P.}},
\bauthor{\bsnm{Di~Cosmo}, \binits{R.}},
\bauthor{\bsnm{Gousios}, \binits{G.}},
\bauthor{\bsnm{Zacchiroli}, \binits{S.}}:
\bctitle{Dependency solving is still hard, but we are getting better at it}.
In: \bbtitle{2020 IEEE 27th International Conference on Software Analysis, Evolution and Reengineering (SANER)},
pp. \bfpage{547}--\blpage{551}
(\byear{2020}).
\bcomment{IEEE}
\end{bchapter}
\endbibitem

%%% 3
\bibitem[\protect\citeauthoryear{Aïdasso et~al.}{2025}]{aïdasso2025buildoptimizationsystematicliterature}
\begin{botherref}
\oauthor{\bsnm{Aïdasso}, \binits{H.}},
\oauthor{\bsnm{Sayagh}, \binits{M.}},
\oauthor{\bsnm{Bordeleau}, \binits{F.}}:
Build Optimization: A Systematic Literature Review
(2025).
\url{https://arxiv.org/abs/2501.11940}
\end{botherref}
\endbibitem

%%% 4
\bibitem[\protect\citeauthoryear{Bogart et~al.}{2021}]{Bogart2021}
\begin{botherref}
\oauthor{\bsnm{Bogart}, \binits{C.}},
\oauthor{\bsnm{K\"{a}stner}, \binits{C.}},
\oauthor{\bsnm{Herbsleb}, \binits{J.}},
\oauthor{\bsnm{Thung}, \binits{F.}}:
{When and How to Make Breaking Changes: Policies and Practices in 18 Open Source Software Ecosystems}.
ACM Trans. Softw. Eng. Methodol.
(2021)
\end{botherref}
\endbibitem

%%% 5
\bibitem[\protect\citeauthoryear{Bifolco et~al.}{2024}]{bifolco2024githubdep}
\begin{bchapter}
\bauthor{\bsnm{Bifolco}, \binits{D.}},
\bauthor{\bsnm{Nocera}, \binits{S.}},
\bauthor{\bsnm{Romano}, \binits{S.}},
\bauthor{\bsnm{Di~Penta}, \binits{M.}},
\bauthor{\bsnm{Francese}, \binits{R.}},
\bauthor{\bsnm{Scanniello}, \binits{G.}}:
\bctitle{On the accuracy of github's dependency graph}.
In: \bbtitle{Proceedings of the 28th International Conference on Evaluation and Assessment in Software Engineering}.
\bsertitle{EASE '24},
pp. \bfpage{242}--\blpage{251}.
\bpublisher{Association for Computing Machinery},
\blocation{New York, NY, USA}
(\byear{2024}).
\doiurl{10.1145/3661167.3661175}
\end{bchapter}
\endbibitem

%%% 6
\bibitem[\protect\citeauthoryear{Bos}{2023}]{bos2023review}
\begin{botherref}
\oauthor{\bsnm{Bos}, \binits{A.M.}}:
A review of attacks against language-based package managers
(2023)
{\href{https://arxiv.org/abs/2302.08959}{{arXiv:2302.08959}}}
\end{botherref}
\endbibitem

%%% 7
\bibitem[\protect\citeauthoryear{Bi et~al.}{2024}]{Bi2024}
\begin{botherref}
\oauthor{\bsnm{Bi}, \binits{T.}},
\oauthor{\bsnm{Xia}, \binits{B.}},
\oauthor{\bsnm{Xing}, \binits{Z.}},
\oauthor{\bsnm{Lu}, \binits{Q.}},
\oauthor{\bsnm{Zhu}, \binits{L.}}:
On the way to sboms: Investigating design issues and solutions in practice.
ACM Trans. Softw. Eng. Methodol.
(2024)
\end{botherref}
\endbibitem

%%% 8
\bibitem[\protect\citeauthoryear{Cass}{2024}]{cass2024spectrum}
\begin{botherref}
\oauthor{\bsnm{Cass}, \binits{S.}}:
The top programming languages 2024
(2024).
\url{https://spectrum.ieee.org/top-programming-languages-2024}
\end{botherref}
\endbibitem

%%% 9
\bibitem[\protect\citeauthoryear{Cleare and Iacob}{2018}]{cleare2018gem}
\begin{bchapter}
\bauthor{\bsnm{Cleare}, \binits{J.}},
\bauthor{\bsnm{Iacob}, \binits{C.}}:
\bctitle{Gemchecker: Reporting on the status of gems in ruby on rails projects}.
In: \bbtitle{2018 IEEE International Conference on Software Maintenance and Evolution (ICSME)},
pp. \bfpage{700}--\blpage{704}
(\byear{2018}).
\doiurl{10.1109/ICSME.2018.00080}
\end{bchapter}
\endbibitem

%%% 10
\bibitem[\protect\citeauthoryear{Cox}{2019}]{Cox19}
\begin{barticle}
\bauthor{\bsnm{Cox}, \binits{R.}}:
\batitle{Surviving software dependencies: Software reuse is finally here but comes with risks.}
\bjtitle{Queue}
\bvolume{17}(\bissue{2}),
\bfpage{24}--\blpage{47}
(\byear{2019})
\doiurl{10.1145/3329781.3344149}
\end{barticle}
\endbibitem

%%% 11
\bibitem[\protect\citeauthoryear{Decan et~al.}{2018}]{decan2018npm}
\begin{bchapter}
\bauthor{\bsnm{Decan}, \binits{A.}},
\bauthor{\bsnm{Mens}, \binits{T.}},
\bauthor{\bsnm{Constantinou}, \binits{E.}}:
\bctitle{On the impact of security vulnerabilities in the npm package dependency network}.
In: \bbtitle{Proceedings of the 15th International Conference on Mining Software Repositories}.
\bsertitle{MSR '18},
pp. \bfpage{181}--\blpage{191}.
\bpublisher{Association for Computing Machinery},
\blocation{New York, NY, USA}
(\byear{2018}).
\doiurl{10.1145/3196398.3196401}
\end{bchapter}
\endbibitem

%%% 12
\bibitem[\protect\citeauthoryear{Goswami et~al.}{2020}]{Goswami2020npmreproducible}
\begin{bchapter}
\bauthor{\bsnm{Goswami}, \binits{P.}},
\bauthor{\bsnm{Gupta}, \binits{S.}},
\bauthor{\bsnm{Li}, \binits{Z.}},
\bauthor{\bsnm{Meng}, \binits{N.}},
\bauthor{\bsnm{Yao}, \binits{D.}}:
\bctitle{Investigating the reproducibility of npm packages}.
In: \bbtitle{2020 IEEE International Conference on Software Maintenance and Evolution (ICSME)},
pp. \bfpage{677}--\blpage{681}
(\byear{2020}).
\doiurl{10.1109/ICSME46990.2020.00071}
\end{bchapter}
\endbibitem

%%% 13
\bibitem[\protect\citeauthoryear{Gu et~al.}{2023}]{gu2023investigating}
\begin{bchapter}
\bauthor{\bsnm{Gu}, \binits{Y.}},
\bauthor{\bsnm{Ying}, \binits{L.}},
\bauthor{\bsnm{Pu}, \binits{Y.}},
\bauthor{\bsnm{Hu}, \binits{X.}},
\bauthor{\bsnm{Chai}, \binits{H.}},
\bauthor{\bsnm{Wang}, \binits{R.}},
\bauthor{\bsnm{Gao}, \binits{X.}},
\bauthor{\bsnm{Duan}, \binits{H.}}:
\bctitle{Investigating package related security threats in software registries}.
In: \bbtitle{2023 IEEE Symposium on Security and Privacy (SP)},
pp. \bfpage{1578}--\blpage{1595}
(\byear{2023}).
\bcomment{IEEE}
\end{bchapter}
\endbibitem

%%% 14
\bibitem[\protect\citeauthoryear{He et~al.}{2025}]{futile2025he}
\begin{botherref}
\oauthor{\bsnm{He}, \binits{H.}},
\oauthor{\bsnm{Vasilescu}, \binits{B.}},
\oauthor{\bsnm{Kästner}, \binits{C.}}:
Pinning is futile: You need more than local dependency versioning to defend against supply chain attacks
(2025).
\url{http://arxiv.org/abs/2502.06662}
\end{botherref}
\endbibitem

%%% 15
\bibitem[\protect\citeauthoryear{Kalliamvakou et~al.}{2014}]{Kalliamvakou2014perils}
\begin{bchapter}
\bauthor{\bsnm{Kalliamvakou}, \binits{E.}},
\bauthor{\bsnm{Gousios}, \binits{G.}},
\bauthor{\bsnm{Blincoe}, \binits{K.}},
\bauthor{\bsnm{Singer}, \binits{L.}},
\bauthor{\bsnm{German}, \binits{D.M.}},
\bauthor{\bsnm{Damian}, \binits{D.}}:
\bctitle{The promises and perils of mining github}.
In: \bbtitle{Proceedings of the 11th Working Conference on Mining Software Repositories}.
\bsertitle{MSR 2014},
pp. \bfpage{92}--\blpage{101}.
\bpublisher{Association for Computing Machinery},
\blocation{New York, NY, USA}
(\byear{2014}).
\doiurl{10.1145/2597073.2597074}
\end{bchapter}
\endbibitem

%%% 16
\bibitem[\protect\citeauthoryear{Krueger}{1992}]{Krueger1992}
\begin{barticle}
\bauthor{\bsnm{Krueger}, \binits{C.W.}}:
\batitle{{Software Reuse}}.
\bjtitle{ACM Computing Surveys}
\bvolume{24}(\bissue{2}),
\bfpage{131}--\blpage{183}
(\byear{1992})
\end{barticle}
\endbibitem

%%% 17
\bibitem[\protect\citeauthoryear{Kalu et~al.}{2024}]{kalu2024industry}
\begin{botherref}
\oauthor{\bsnm{Kalu}, \binits{K.G.}},
\oauthor{\bsnm{Singla}, \binits{T.}},
\oauthor{\bsnm{Okafor}, \binits{C.}},
\oauthor{\bsnm{Torres-Arias}, \binits{S.}},
\oauthor{\bsnm{Davis}, \binits{J.C.}}:
An industry interview study of software signing for supply chain security.
arXiv preprint arXiv:2406.08198
(2024)
\end{botherref}
\endbibitem

%%% 18
\bibitem[\protect\citeauthoryear{Kabir et~al.}{2022}]{kabir2022securitybest}
\begin{bchapter}
\bauthor{\bsnm{Kabir}, \binits{M.M.A.}},
\bauthor{\bsnm{Wang}, \binits{Y.}},
\bauthor{\bsnm{Yao}, \binits{D.}},
\bauthor{\bsnm{Meng}, \binits{N.}}:
\bctitle{How do developers follow security-relevant best practices when using npm packages?}
In: \bbtitle{2022 IEEE Secure Development Conference (SecDev)},
pp. \bfpage{77}--\blpage{83}
(\byear{2022}).
\doiurl{10.1109/SecDev53368.2022.00027}
\end{bchapter}
\endbibitem

%%% 19
\bibitem[\protect\citeauthoryear{Li et~al.}{2023}]{Hao23}
\begin{barticle}
\bauthor{\bsnm{Li}, \binits{H.}},
\bauthor{\bsnm{C{\^{o}}go}, \binits{F.R.}},
\bauthor{\bsnm{Bezemer}, \binits{C.}}:
\batitle{An empirical study of yanked releases in the rust package registry}.
\bjtitle{{IEEE} Trans. Software Eng.}
\bvolume{49}(\bissue{1}),
\bfpage{437}--\blpage{449}
(\byear{2023})
\end{barticle}
\endbibitem

%%% 20
\bibitem[\protect\citeauthoryear{Liu et~al.}{2022}]{liu2022demystifynpm}
\begin{bchapter}
\bauthor{\bsnm{Liu}, \binits{C.}},
\bauthor{\bsnm{Chen}, \binits{S.}},
\bauthor{\bsnm{Fan}, \binits{L.}},
\bauthor{\bsnm{Chen}, \binits{B.}},
\bauthor{\bsnm{Liu}, \binits{Y.}},
\bauthor{\bsnm{Peng}, \binits{X.}}:
\bctitle{Demystifying the vulnerability propagation and its evolution via dependency trees in the npm ecosystem}.
In: \bbtitle{Proceedings of the 44th International Conference on Software Engineering}.
\bsertitle{ICSE '22},
pp. \bfpage{672}--\blpage{684}.
\bpublisher{Association for Computing Machinery},
\blocation{New York, NY, USA}
(\byear{2022}).
\doiurl{10.1145/3510003.3510142}
\end{bchapter}
\endbibitem

%%% 21
\bibitem[\protect\citeauthoryear{Ladisa et~al.}{2023}]{ladisa2023sok}
\begin{bchapter}
\bauthor{\bsnm{Ladisa}, \binits{P.}},
\bauthor{\bsnm{Plate}, \binits{H.}},
\bauthor{\bsnm{Martinez}, \binits{M.}},
\bauthor{\bsnm{Barais}, \binits{O.}}:
\bctitle{Sok: Taxonomy of attacks on open-source software supply chains}.
In: \bbtitle{2023 IEEE Symposium on Security and Privacy (SP)},
pp. \bfpage{1509}--\blpage{1526}
(\byear{2023}).
\bcomment{IEEE}
\end{bchapter}
\endbibitem

%%% 22
\bibitem[\protect\citeauthoryear{Lamb and Zacchiroli}{2022}]{Lamb2022}
\begin{barticle}
\bauthor{\bsnm{Lamb}, \binits{C.}},
\bauthor{\bsnm{Zacchiroli}, \binits{S.}}:
\batitle{Reproducible builds: Increasing the integrity of software supply chains}.
\bjtitle{IEEE Software}
\bvolume{39}(\bissue{2}),
\bfpage{62}--\blpage{70}
(\byear{2022})
\end{barticle}
\endbibitem

%%% 23
\bibitem[\protect\citeauthoryear{Mohayeji et~al.}{2023}]{mohayeji2023dependabot}
\begin{bchapter}
\bauthor{\bsnm{Mohayeji}, \binits{H.}},
\bauthor{\bsnm{Agaronian}, \binits{A.}},
\bauthor{\bsnm{Constantinou}, \binits{E.}},
\bauthor{\bsnm{Zannone}, \binits{N.}},
\bauthor{\bsnm{Serebrenik}, \binits{A.}}:
\bctitle{Investigating the resolution of vulnerable dependencies with dependabot security updates}.
In: \bbtitle{2023 IEEE/ACM 20th International Conference on Mining Software Repositories (MSR)},
pp. \bfpage{234}--\blpage{246}
(\byear{2023}).
\doiurl{10.1109/MSR59073.2023.00042}
\end{bchapter}
\endbibitem

%%% 24
\bibitem[\protect\citeauthoryear{Mohayeji et~al.}{2025}]{Mohayeji2025dependabot}
\begin{botherref}
\oauthor{\bsnm{Mohayeji}, \binits{H.}},
\oauthor{\bsnm{Agaronian}, \binits{A.}},
\oauthor{\bsnm{Constantinou}, \binits{E.}},
\oauthor{\bsnm{Zannone}, \binits{N.}},
\oauthor{\bsnm{Serebrenik}, \binits{A.}}:
Securing dependencies: A comprehensive study of dependabot’s impact on vulnerability mitigation.
Empirical Software Engineering
\textbf{30}(3)
(2025)
\doiurl{10.1007/s10664-025-10638-w}
\end{botherref}
\endbibitem

%%% 25
\bibitem[\protect\citeauthoryear{Mohagheghi and Conradi}{2007}]{mohagheghi2007quality}
\begin{barticle}
\bauthor{\bsnm{Mohagheghi}, \binits{P.}},
\bauthor{\bsnm{Conradi}, \binits{R.}}:
\batitle{Quality, productivity and economic benefits of software reuse: a review of industrial studies}.
\bjtitle{Empirical Software Engineering}
\bvolume{12},
\bfpage{471}--\blpage{516}
(\byear{2007})
\end{barticle}
\endbibitem

%%% 26
\bibitem[\protect\citeauthoryear{Malterud et~al.}{2016}]{Malterud2016power}
\begin{barticle}
\bauthor{\bsnm{Malterud}, \binits{K.}},
\bauthor{\bsnm{Siersma}, \binits{V.D.}},
\bauthor{\bsnm{Guassora}, \binits{A.D.}}:
\batitle{Sample size in qualitative interview studies: Guided by information power}.
\bjtitle{Qualitative health research}
\bvolume{26}(\bissue{13}),
\bfpage{1753}--\blpage{1760}
(\byear{2016})
\doiurl{10.1177/1049732315617444}
\end{barticle}
\endbibitem

%%% 27
\bibitem[\protect\citeauthoryear{Ochoa et~al.}{2022}]{ochoa2021bad}
\begin{botherref}
\oauthor{\bsnm{Ochoa}, \binits{L.}},
\oauthor{\bsnm{Degueule}, \binits{T.}},
\oauthor{\bsnm{Falleri}, \binits{J.-R.}},
\oauthor{\bsnm{Vinju}, \binits{J.}}:
Breaking bad? semantic versioning and impact of breaking changes in maven central: An external and differentiated replication study.
Empirical Softw. Engg.
\textbf{27}(3)
(2022)
\doiurl{10.1007/s10664-021-10052-y}
\end{botherref}
\endbibitem

%%% 28
\bibitem[\protect\citeauthoryear{Pinckney et~al.}{2023}]{pinckney2023maxsmt}
\begin{bchapter}
\bauthor{\bsnm{Pinckney}, \binits{D.}},
\bauthor{\bsnm{Cassano}, \binits{F.}},
\bauthor{\bsnm{Guha}, \binits{A.}},
\bauthor{\bsnm{Bell}, \binits{J.}},
\bauthor{\bsnm{Culpo}, \binits{M.}},
\bauthor{\bsnm{Gamblin}, \binits{T.}}:
\bctitle{Flexible and optimal dependency management via max-smt}.
In: \bbtitle{2023 IEEE/ACM 45th International Conference on Software Engineering (ICSE)},
pp. \bfpage{1418}--\blpage{1429}
(\byear{2023}).
\doiurl{10.1109/ICSE48619.2023.00124}
\end{bchapter}
\endbibitem

%%% 29
\bibitem[\protect\citeauthoryear{Patra et~al.}{2018}]{Patra2018conflictjs}
\begin{bchapter}
\bauthor{\bsnm{Patra}, \binits{J.}},
\bauthor{\bsnm{Dixit}, \binits{P.N.}},
\bauthor{\bsnm{Pradel}, \binits{M.}}:
\bctitle{Conflictjs: Finding and understanding conflicts between javascript libraries}.
In: \bbtitle{2018 IEEE/ACM 40th International Conference on Software Engineering (ICSE)},
pp. \bfpage{741}--\blpage{751}
(\byear{2018}).
\doiurl{10.1145/3180155.3180184}
\end{bchapter}
\endbibitem

%%% 30
\bibitem[\protect\citeauthoryear{Reyes et~al.}{2024}]{frank2024bg}
\begin{bchapter}
\bauthor{\bsnm{Reyes}, \binits{F.}},
\bauthor{\bsnm{Baudry}, \binits{B.}},
\bauthor{\bsnm{Monperrus}, \binits{M.}}:
\bctitle{{ Breaking-Good: Explaining Breaking Dependency Updates with Build Analysis }}.
In: \bbtitle{2024 IEEE International Conference on Source Code Analysis and Manipulation (SCAM)},
pp. \bfpage{36}--\blpage{46}.
\bpublisher{IEEE Computer Society},
\blocation{Los Alamitos, CA, USA}
(\byear{2024}).
\doiurl{10.1109/SCAM63643.2024.00014}
\end{bchapter}
\endbibitem

%%% 31
\bibitem[\protect\citeauthoryear{Rombaut et~al.}{2023}]{Rombaut2023}
\begin{botherref}
\oauthor{\bsnm{Rombaut}, \binits{B.}},
\oauthor{\bsnm{Cogo}, \binits{F.R.}},
\oauthor{\bsnm{Adams}, \binits{B.}},
\oauthor{\bsnm{Hassan}, \binits{A.E.}}:
There’s no such thing as a free lunch: Lessons learned from exploring the overhead introduced by the greenkeeper dependency bot in npm.
ACM Trans. Softw. Eng. Methodol.
(2023)
\end{botherref}
\endbibitem

%%% 32
\bibitem[\protect\citeauthoryear{Reyes et~al.}{2024}]{frank2924bump}
\begin{bchapter}
\bauthor{\bsnm{Reyes}, \binits{F.}},
\bauthor{\bsnm{Gamage}, \binits{Y.}},
\bauthor{\bsnm{Skoglund}, \binits{G.}},
\bauthor{\bsnm{Baudry}, \binits{B.}},
\bauthor{\bsnm{Monperrus}, \binits{M.}}:
\bctitle{Bump: A benchmark of reproducible breaking dependency updates}.
In: \bbtitle{2024 IEEE International Conference on Software Analysis, Evolution and Reengineering (SANER)},
pp. \bfpage{159}--\blpage{170}
(\byear{2024}).
\doiurl{10.1109/SANER60148.2024.00024}
\end{bchapter}
\endbibitem

%%% 33
\bibitem[\protect\citeauthoryear{Rausch et~al.}{2017}]{rausch2017empirical}
\begin{bchapter}
\bauthor{\bsnm{Rausch}, \binits{T.}},
\bauthor{\bsnm{Hummer}, \binits{W.}},
\bauthor{\bsnm{Leitner}, \binits{P.}},
\bauthor{\bsnm{Schulte}, \binits{S.}}:
\bctitle{An empirical analysis of build failures in the continuous integration workflows of java-based open-source software}.
In: \bbtitle{2017 IEEE/ACM 14th International Conference on Mining Software Repositories (MSR)},
pp. \bfpage{345}--\blpage{355}
(\byear{2017}).
\bcomment{IEEE}
\end{bchapter}
\endbibitem

%%% 34
\bibitem[\protect\citeauthoryear{Radford et~al.}{2023}]{radford2023whisper}
\begin{bchapter}
\bauthor{\bsnm{Radford}, \binits{A.}},
\bauthor{\bsnm{Kim}, \binits{J.W.}},
\bauthor{\bsnm{Xu}, \binits{T.}},
\bauthor{\bsnm{Brockman}, \binits{G.}},
\bauthor{\bsnm{McLeavey}, \binits{C.}},
\bauthor{\bsnm{Sutskever}, \binits{I.}}:
\bctitle{Robust speech recognition via large-scale weak supervision}.
In: \bbtitle{Proceedings of the 40th International Conference on Machine Learning}.
\bsertitle{ICML'23}
(\byear{2023})
\end{bchapter}
\endbibitem

%%% 35
\bibitem[\protect\citeauthoryear{Rahman et~al.}{2025}]{rahman2025pinning}
\begin{botherref}
\oauthor{\bsnm{Rahman}, \binits{I.}},
\oauthor{\bsnm{Marley}, \binits{J.}},
\oauthor{\bsnm{Enck}, \binits{W.}},
\oauthor{\bsnm{Williams}, \binits{L.}}:
Which Is Better For Reducing Outdated and Vulnerable Dependencies: Pinning or Floating?
(2025).
\url{https://arxiv.org/abs/2510.08609}
\end{botherref}
\endbibitem

%%% 36
\bibitem[\protect\citeauthoryear{Rahman et~al.}{2025}]{rahman2025vulnerabilitydataproblempredicting}
\begin{botherref}
\oauthor{\bsnm{Rahman}, \binits{I.}},
\oauthor{\bsnm{Paramitha}, \binits{R.}},
\oauthor{\bsnm{Zahan}, \binits{N.}},
\oauthor{\bsnm{Magill}, \binits{S.}},
\oauthor{\bsnm{Enck}, \binits{W.}},
\oauthor{\bsnm{Williams}, \binits{L.}}:
No Vulnerability Data, No Problem: Towards Predicting Mean Time To Remediate In Open Source Software Dependencies
(2025).
\url{https://arxiv.org/abs/2403.17382}
\end{botherref}
\endbibitem

%%% 37
\bibitem[\protect\citeauthoryear{Schorlemmer et~al.}{2024}]{schorlemmer2024signing}
\begin{bchapter}
\bauthor{\bsnm{Schorlemmer}, \binits{T.R.}},
\bauthor{\bsnm{Kalu}, \binits{K.G.}},
\bauthor{\bsnm{Chigges}, \binits{L.}},
\bauthor{\bsnm{Ko}, \binits{K.M.}},
\bauthor{\bsnm{Ishgair}, \binits{E.A.}},
\bauthor{\bsnm{Bagchi}, \binits{S.}},
\bauthor{\bsnm{Torres-Arias}, \binits{S.}},
\bauthor{\bsnm{Davis}, \binits{J.C.}}:
\bctitle{{ Signing in Four Public Software Package Registries: Quantity, Quality, and Influencing Factors }}.
In: \bbtitle{2024 IEEE Symposium on Security and Privacy (SP)},
pp. \bfpage{1160}--\blpage{1178}.
\bpublisher{IEEE Computer Society},
\blocation{Los Alamitos, CA, USA}
(\byear{2024}).
\doiurl{10.1109/SP54263.2024.00215}
\end{bchapter}
\endbibitem

%%% 38
\bibitem[\protect\citeauthoryear{Spinellis}{2012}]{spinellis2012package}
\begin{barticle}
\bauthor{\bsnm{Spinellis}, \binits{D.}}:
\batitle{Package management systems}.
\bjtitle{IEEE software}
\bvolume{29}(\bissue{2}),
\bfpage{84}--\blpage{86}
(\byear{2012})
\end{barticle}
\endbibitem

%%% 39
\bibitem[\protect\citeauthoryear{Schueller et~al.}{2022}]{Schueller2022rust}
\begin{barticle}
\bauthor{\bsnm{Schueller}, \binits{W.}},
\bauthor{\bsnm{Wachs}, \binits{J.}},
\bauthor{\bsnm{Servedio}, \binits{V.D.P.}},
\bauthor{\bsnm{Thurner}, \binits{S.}},
\bauthor{\bsnm{Loreto}, \binits{V.}}:
\batitle{Evolving collaboration, dependencies, and use in the rust open source software ecosystem}.
\bjtitle{Scientific data}
\bvolume{9}(\bissue{1}),
\bfpage{703}
(\byear{2022})
\doiurl{10.1038/s41597-022-01819-z}
\end{barticle}
\endbibitem

%%% 40
\bibitem[\protect\citeauthoryear{Vaidya et~al.}{2021}]{vaidya2019security}
\begin{botherref}
\oauthor{\bsnm{Vaidya}, \binits{R.K.}},
\oauthor{\bsnm{Carli}, \binits{L.D.}},
\oauthor{\bsnm{Davidson}, \binits{D.}},
\oauthor{\bsnm{Rastogi}, \binits{V.}}:
Security issues in language-based software ecosystems
(2021)
{\href{https://arxiv.org/abs/1903.02613}{{arXiv:1903.02613}}}
\end{botherref}
\endbibitem

%%% 41
\bibitem[\protect\citeauthoryear{Venturini et~al.}{2023}]{venturini2023dependubreaknpm}
\begin{botherref}
\oauthor{\bsnm{Venturini}, \binits{D.}},
\oauthor{\bsnm{Cogo}, \binits{F.R.}},
\oauthor{\bsnm{Polato}, \binits{I.}},
\oauthor{\bsnm{Gerosa}, \binits{M.A.}},
\oauthor{\bsnm{Wiese}, \binits{I.S.}}:
I depended on you and you broke me: An empirical study of manifesting breaking changes in client packages.
ACM Trans. Softw. Eng. Methodol.
(2023)
\end{botherref}
\endbibitem

%%% 42
\bibitem[\protect\citeauthoryear{Williams et~al.}{2025}]{williams25directions}
\begin{botherref}
\oauthor{\bsnm{Williams}, \binits{L.}},
\oauthor{\bsnm{Benedetti}, \binits{G.}},
\oauthor{\bsnm{Hamer}, \binits{S.}},
\oauthor{\bsnm{Paramitha}, \binits{R.}},
\oauthor{\bsnm{Rahman}, \binits{I.}},
\oauthor{\bsnm{Tamanna}, \binits{M.}},
\oauthor{\bsnm{Tystahl}, \binits{G.}},
\oauthor{\bsnm{Zahan}, \binits{N.}},
\oauthor{\bsnm{Morrison}, \binits{P.}},
\oauthor{\bsnm{Acar}, \binits{Y.}},
\oauthor{\bsnm{Cukier}, \binits{M.}},
\oauthor{\bsnm{K\"{a}stner}, \binits{C.}},
\oauthor{\bsnm{Kapravelos}, \binits{A.}},
\oauthor{\bsnm{Wermke}, \binits{D.}},
\oauthor{\bsnm{Enck}, \binits{W.}}:
Research directions in software supply chain security.
ACM Trans. Softw. Eng. Methodol.
(2025)
\end{botherref}
\endbibitem

%%% 43
\bibitem[\protect\citeauthoryear{Wang et~al.}{2021}]{wang2021gopath}
\begin{bchapter}
\bauthor{\bsnm{Wang}, \binits{Y.}},
\bauthor{\bsnm{Qiao}, \binits{L.}},
\bauthor{\bsnm{Xu}, \binits{C.}},
\bauthor{\bsnm{Liu}, \binits{Y.}},
\bauthor{\bsnm{Cheung}, \binits{S.-C.}},
\bauthor{\bsnm{Meng}, \binits{N.}},
\bauthor{\bsnm{Yu}, \binits{H.}},
\bauthor{\bsnm{Zhu}, \binits{Z.}}:
\bctitle{Hero: On the chaos when path meets modules}.
In: \bbtitle{Proceedings of the 43rd International Conference on Software Engineering}.
\bsertitle{ICSE '21},
pp. \bfpage{99}--\blpage{111}
(\byear{2021}).
\doiurl{10.1109/ICSE43902.2021.00022}
\end{bchapter}
\endbibitem

%%% 44
\bibitem[\protect\citeauthoryear{Wang et~al.}{2020}]{wang2020watchman}
\begin{bchapter}
\bauthor{\bsnm{Wang}, \binits{Y.}},
\bauthor{\bsnm{Wen}, \binits{M.}},
\bauthor{\bsnm{Liu}, \binits{Y.}},
\bauthor{\bsnm{Wang}, \binits{Y.}},
\bauthor{\bsnm{Li}, \binits{Z.}},
\bauthor{\bsnm{Wang}, \binits{C.}},
\bauthor{\bsnm{Yu}, \binits{H.}},
\bauthor{\bsnm{Cheung}, \binits{S.-C.}},
\bauthor{\bsnm{Xu}, \binits{C.}},
\bauthor{\bsnm{Zhu}, \binits{Z.}}:
\bctitle{Watchman: Monitoring dependency conflicts for python library ecosystem}.
In: \bbtitle{2020 IEEE/ACM 42nd International Conference on Software Engineering (ICSE)},
pp. \bfpage{125}--\blpage{135}
(\byear{2020}).
\doiurl{10.1145/3377811.3380426}
\end{bchapter}
\endbibitem

%%% 45
\bibitem[\protect\citeauthoryear{Wang et~al.}{2023}]{wang2022smartpip}
\begin{bchapter}
\bauthor{\bsnm{Wang}, \binits{C.}},
\bauthor{\bsnm{Wu}, \binits{R.}},
\bauthor{\bsnm{Song}, \binits{H.}},
\bauthor{\bsnm{Shu}, \binits{J.}},
\bauthor{\bsnm{Li}, \binits{G.}}:
\bctitle{smartpip: A smart approach to resolving python dependency conflict issues}.
In: \bbtitle{Proceedings of the 37th IEEE/ACM International Conference on Automated Software Engineering}.
\bsertitle{ASE '22}.
\bpublisher{Association for Computing Machinery},
\blocation{New York, NY, USA}
(\byear{2023}).
\doiurl{10.1145/3551349.3560437}
\end{bchapter}
\endbibitem

%%% 46
\bibitem[\protect\citeauthoryear{Wang et~al.}{2025}]{wang2025peer}
\begin{bchapter}
\bauthor{\bsnm{Wang}, \binits{X.}},
\bauthor{\bsnm{Wang}, \binits{M.}},
\bauthor{\bsnm{Shen}, \binits{W.}},
\bauthor{\bsnm{Chang}, \binits{R.}}:
\bctitle{{ Understanding and Detecting Peer Dependency Resolving Loop in npm Ecosystem }}.
In: \bbtitle{2025 IEEE/ACM 47th International Conference on Software Engineering (ICSE)},
pp. \bfpage{591}--\blpage{591}.
\bpublisher{IEEE Computer Society},
\blocation{Los Alamitos, CA, USA}
(\byear{2025}).
\doiurl{10.1109/ICSE55347.2025.00054}
\end{bchapter}
\endbibitem

%%% 47
\bibitem[\protect\citeauthoryear{Wang et~al.}{2019}]{wang2019stacktrace}
\begin{bchapter}
\bauthor{\bsnm{Wang}, \binits{Y.}},
\bauthor{\bsnm{Wen}, \binits{M.}},
\bauthor{\bsnm{Wu}, \binits{R.}},
\bauthor{\bsnm{Liu}, \binits{Z.}},
\bauthor{\bsnm{Tan}, \binits{S.H.}},
\bauthor{\bsnm{Zhu}, \binits{Z.}},
\bauthor{\bsnm{Yu}, \binits{H.}},
\bauthor{\bsnm{Cheung}, \binits{S.-C.}}:
\bctitle{Could i have a stack trace to examine the dependency conflict issue?}
In: \bbtitle{2019 IEEE/ACM 41st International Conference on Software Engineering (ICSE)},
pp. \bfpage{572}--\blpage{583}
(\byear{2019}).
\doiurl{10.1109/ICSE.2019.00068}
\end{bchapter}
\endbibitem

%%% 48
\bibitem[\protect\citeauthoryear{Yu et~al.}{2024}]{yu2024sbomgen}
\begin{bchapter}
\bauthor{\bsnm{Yu}, \binits{S.}},
\bauthor{\bsnm{Song}, \binits{W.}},
\bauthor{\bsnm{Hu}, \binits{X.}},
\bauthor{\bsnm{Yin}, \binits{H.}}:
\bctitle{On the correctness of metadata-based sbom generation: A differential analysis approach}.
In: \bbtitle{2024 54th Annual IEEE/IFIP International Conference on Dependable Systems and Networks (DSN)},
pp. \bfpage{29}--\blpage{36}
(\byear{2024}).
\doiurl{10.1109/DSN58291.2024.00018}
\end{bchapter}
\endbibitem

\end{thebibliography}

\end{document}